\documentclass{article} 
\usepackage{iclr2027_conference,times}

\usepackage{amsmath,amsfonts,bm}

\def\eqref#1{equation~\ref{#1}}

\def\1{\bm{1}}

\DeclareMathAlphabet{\mathsfit}{\encodingdefault}{\sfdefault}{m}{sl}
\SetMathAlphabet{\mathsfit}{bold}{\encodingdefault}{\sfdefault}{bx}{n}

\usepackage{hyperref}
\usepackage{url}
\usepackage{graphicx}
\usepackage{amssymb}
\usepackage{booktabs}
\usepackage{multirow}
\usepackage{xcolor}
\usepackage{colortbl}
\usepackage{pifont}
\usepackage{float}
\usepackage{wrapfig}
\usepackage{placeins}
\usepackage{afterpage}
\usepackage{algorithm}
\usepackage{algorithmic}
\usepackage[most]{tcolorbox}

\newtcblisting{promptbox}[2][]{%
  enhanced,
  breakable,
  listing only,
  listing file=\jobname-listing-\thetcblistingcount.listing,
  colback=black!2,
  colframe=black!60,
  colbacktitle=black!65,
  coltitle=white,
  title={#2},
  fonttitle=\bfseries\large,
  listing options={%
    basicstyle=\ttfamily\footnotesize,
    columns=fullflexible,
    keepspaces=true,
    breaklines=true,
    breakatwhitespace=true,
    breakautoindent=false,
    breakindent=0pt,
    showstringspaces=false
  },
  boxrule=0.7pt,
  arc=3mm,
  outer arc=3mm,
  left=7pt,
  right=7pt,
  top=5pt,
  bottom=7pt,
  #1
}

\definecolor{unimasoaccent}{HTML}{0072B2}
\definecolor{unimasorow}{RGB}{224,242,250}
\definecolor{checkgreen}{RGB}{0,145,75}
\definecolor{crossred}{RGB}{220,45,45}
\newcommand{\cmark}{\textcolor{checkgreen}{\ding{51}}}
\newcommand{\xmark}{\textcolor{crossred}{\ding{55}}}

\title{Not Every Agent Needs the Strongest Model: Unified Prompt--Model Optimization for Heterogeneous LLM-based Multi-Agent Systems}

\author{\parbox[t]{\textwidth}{%
\raggedright
\fontsize{10}{12}\selectfont\bfseries
Mingxiu~Cai, Haoyang~Xian, Xinyu~Ma, Yutong~Wang,
Zhenxi~Song, Jing~Li, and Xuebo~Liu\thanks{Corresponding author.}\\
{\normalfont\small Institute of Computing and Intelligence, Harbin Institute of Technology, Shenzhen}\\
{\normalfont\scriptsize
\texttt{\{26b951113, 2023311323, 26b351020, 25b351016\}@stu.hit.edu.cn}\\
\texttt{\{songzhenxi, li.jing, liuxuebo\}@hit.edu.cn}\\
}%
}}

\iclrfinalcopy
\begin{document}

\maketitle
\lhead{Unified Prompt--Model Optimization for Heterogeneous LLM-based Multi-Agent Systems}
\vspace{-0.18in}

\begin{abstract}
Large language model (LLM)-based multi-agent systems (MAS) rely on role-specific
prompts and execution models to solve complex tasks collaboratively. However,
prompt optimization typically keeps model assignments fixed, leaving differences
in role requirements and their associated inference costs outside the search.
We present \textbf{\textsc{UniMASO}}, a \textbf{Uni}fied framework for
heterogeneous \textbf{M}ulti-\textbf{A}gent \textbf{S}ystem
\textbf{O}ptimization through the joint adaptation of prompts and model
assignments. UniMASO evaluates candidate updates through
local, downstream, and global feedback, using cost-aware Pareto search to retain
useful quality--cost trade-offs. Topology-aware co-evolution then coordinates
these updates according to agent roles and dependencies. Cross-level trajectory
memory further supplies historical successes and failures to guide subsequent
optimization. Experiments on five benchmarks and five execution topologies show
that UniMASO achieves competitive task performance at substantially lower
deployment cost than the strongest homogeneous configurations. Under centralized
execution, it improves the average score from 69.79 to 70.40 while reducing
token-weighted deployment cost by 85.2\% relative to DeepSeek-V4-Pro with MASPO.
These results support joint prompt--model adaptation as an effective approach to
efficient multi-agent collaboration.
The code is available at \url{https://github.com/Achilles008/UniMASO_main}.
\end{abstract}

\section{Introduction}
\label{sec:introduction}
\enlargethispage{\baselineskip}

LLM-based multi-agent systems (MAS) decompose complex tasks through role specialization and communication \citep{li2023camel,wu2024autogen,chen2024agentverse,hong2024metagpt,zhuge2024gptswarm,wang-etal-2025-agentdropout,tian-etal-2025-agentinit}. Each agent's prompt shapes its behavior and the information passed downstream, while its execution model determines capabilities and inference cost. System quality depends on how both choices align across the communication topology (Table~\ref{tab:prompt_model_comparison}).

\begin{wraptable}[12]{r}{0.45\textwidth}
    \vspace{-1.0\baselineskip}
    \centering
    \begin{minipage}{0.92\linewidth}
        \caption{Comparison of prompt and model optimization paradigms.}
        \label{tab:prompt_model_comparison}
    \end{minipage}
    \scriptsize
    \setlength{\tabcolsep}{3.4pt}
    \renewcommand{\arraystretch}{1.08}
    \begin{tabular}{@{}l@{\hspace{0.65em}}cc@{}}
        \toprule
        \textbf{Method} & \textbf{Prompt} & \textbf{Model} \\
        \midrule
        MIPRO~\citep{opsahlong2024optimizing} & \cmark & \xmark \\
        GPTSwarm~\citep{zhuge2024gptswarm} & \cmark & \xmark \\
        MASS~\citep{zhou2026multiagent} & \cmark & \xmark \\
        MASPOB~\citep{hong2026maspob} & \cmark & \xmark \\
        MASPO~\citep{wang2026maspo} & \cmark & \xmark \\
        \addlinespace[1pt]
        MasRouter~\citep{yue2025masrouter} & \xmark & \cmark \\
        Hetero-Designer~\citep{zhang2026heterodesigner} & \xmark & \cmark \\
        \midrule
        \rowcolor{unimasorow}
        \textbf{\textsc{UniMASO} (Ours)} & \cmark & \cmark \\
        \bottomrule
    \end{tabular}
\end{wraptable}

Automatic prompt optimization refines instructions from task feedback, and recent MAS methods coordinate role prompts \citep{yang2024large,fernando2024promptbreeder,opsahlong2024optimizing,cheng2024trace,zhou2026multiagent,zhang2026mapro,wang2026maspo,hong2026maspob}; they generally fix execution models. Conversely, model routing and heterogeneous agent design assign models by task or role \citep{chen2024frugalgpt,ong2025routellm,chen2025optimizing,yue2025masrouter,zhang2026heterodesigner}, but typically fix prompts, leaving behavior outside model search. Fixed model assignments create \emph{role--capability misalignment}: they can waste strong models on routine roles or bottleneck demanding roles with lightweight models, while prompt edits cannot supply missing capabilities.

Changes to either choice propagate through the agent graph, altering downstream context, end-to-end quality, and cost; independent search can favor locally attractive but system-level poor configurations. This coupling motivates joint prompt--model search across the full execution topology.

\begin{wrapfigure}[12]{r}{0.49\textwidth}
    \centering
    \includegraphics[width=\linewidth]{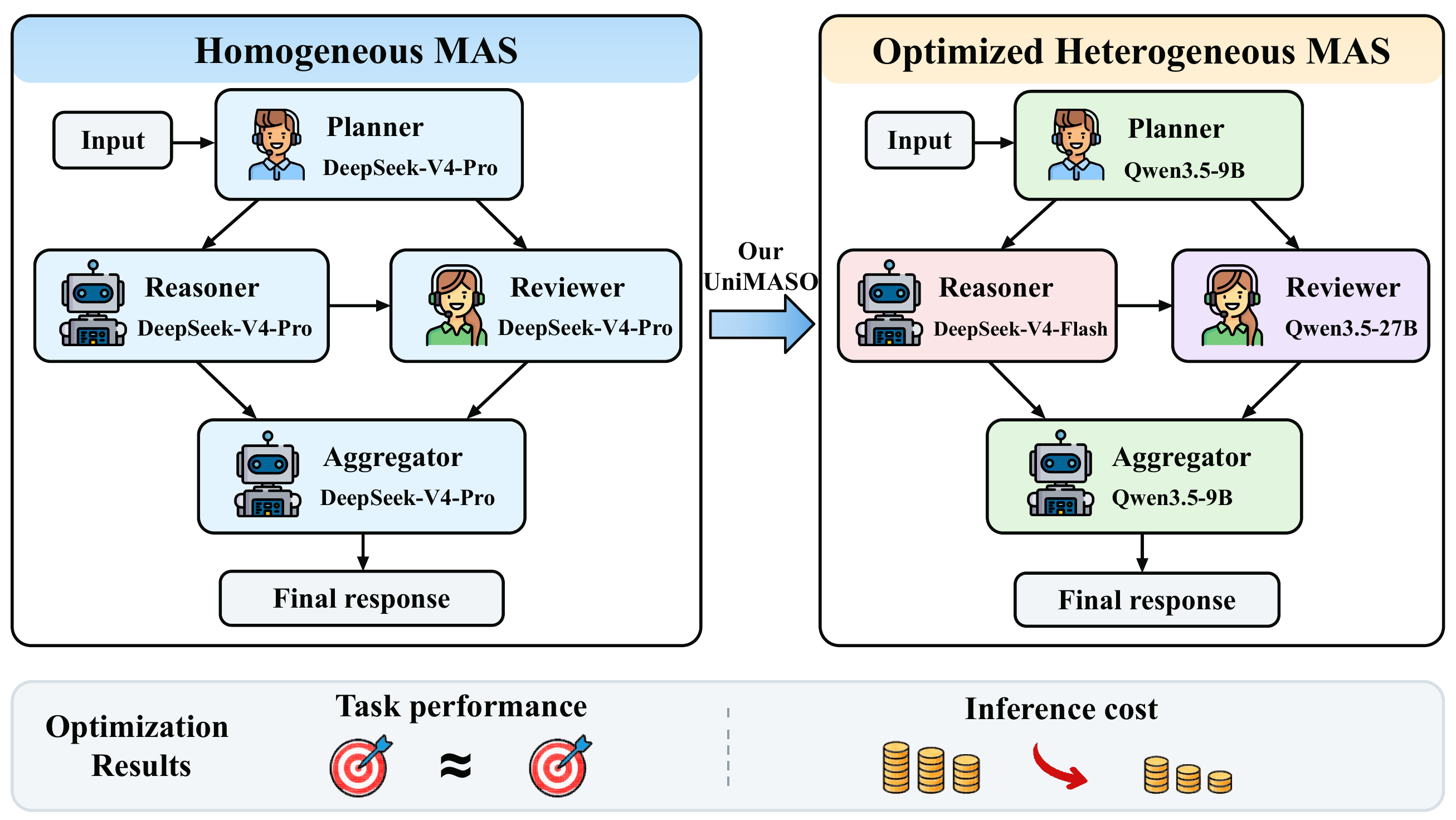}
    \vspace{-12pt}
    \setlength{\abovecaptionskip}{0pt}
        \caption{\footnotesize Homogeneous vs. Heterogeneous MAS.}
    \label{fig:homogeneous_heterogeneous}
\end{wrapfigure}

To address this gap, we introduce \textbf{UniMASO}, which jointly optimizes agent prompts and model assignments in heterogeneous MAS (Figure~\ref{fig:homogeneous_heterogeneous}). It evaluates candidate updates through local, downstream, and global feedback, then applies cost-aware Pareto search to retain quality--cost trade-offs. Topology-aware co-evolution coordinates updates across dependent agents, while cross-level trajectory memory guides subsequent search.
Across five benchmarks and five execution topologies, we assess quality--cost trade-offs, topology generalization, and the contributions of trajectory memory and search controls. Under centralized execution, UniMASO raises the average score from 69.79 to 70.40 while reducing token-weighted deployment cost by 85.2\% relative to DeepSeek-V4-Pro with MASPO.

Our contributions are summarized as follows:
\begin{itemize}
    \item We formulate heterogeneous MAS optimization over prompt--model
    configurations, making role--capability misalignment explicit.
    \item We develop cost- and topology-aware search that couples local
    multi-granularity evaluation with system-level co-evolution.
    \item We introduce cross-level trajectory memory that turns local feedback,
    global outcomes, and cost observations into experience for later rounds.
\end{itemize}

\section{Related Work}
\label{sec:related_work}

\subsection{Multi-Agent Systems and Prompt Optimization}

LLM-based multi-agent systems coordinate model calls through specialized roles
and message exchange. CAMEL, AutoGen, AgentVerse, and MetaGPT establish role-based
workflows \citep{li2023camel,wu2024autogen,chen2024agentverse,hong2024metagpt},
while GPTSwarm and MASS automate system construction by optimizing agent graphs or
co-evolving prompts and topologies \citep{zhuge2024gptswarm,zhou2026multiagent}.
Prompt optimizers search natural-language instructions through candidate scores,
evolutionary search, execution traces, or language feedback
\citep{yang2024large,fernando2024promptbreeder,cheng2024trace,agrawal2026gepa}.
MIPRO jointly optimizes instructions and demonstrations in modular programs
\citep{opsahlong2024optimizing}. Because upstream outputs shape successor context,
local prompt gains need not improve final responses. MAPRO coordinates node-wise candidates with probabilistic inference \citep{zhang2026mapro},
whereas MASPO evaluates local, downstream, and global effects through topology-guided
search \citep{wang2026maspo}.
MASPOB allocates optimization budgets using graph
representations and bandit search \citep{hong2026maspob}. These methods keep
node-level execution models fixed. UniMASO jointly searches prompts and model
assignments, using cross-round trajectories to balance system utility and
inference cost.
\subsection{Heterogeneous Model Assignment}

Model capability and invocation cost motivate cost-aware model selection. FrugalGPT
reduces inference cost through learned cascades, while RouteLLM routes queries
between models using preference data \citep{chen2024frugalgpt,ong2025routellm}.
Compound language-model systems optimize model combinations under performance and
budget constraints \citep{chen2025optimizing}. In multi-agent settings, MasRouter
selects execution models based on task and agent characteristics, while
Hetero-Designer jointly assigns roles, models, and collaboration structures
\citep{yue2025masrouter,zhang2026heterodesigner}. These methods route models under
fixed prompts or configure each query independently, decoupling model choice from
the evolving prompt context and its downstream execution effects. UniMASO
integrates heterogeneous assignment into iterative prompt optimization, making
model choices depend on each candidate prompt, its topological position, downstream
influence, and realized inference cost. This coupling matches model capacity to
role requirements while optimizing the system-level quality--cost objective.
\section{Methodology}
\label{sec:method}

\begin{figure*}[t]
    \centering
    \setlength{\abovecaptionskip}{1pt}
    \includegraphics[width=\textwidth]{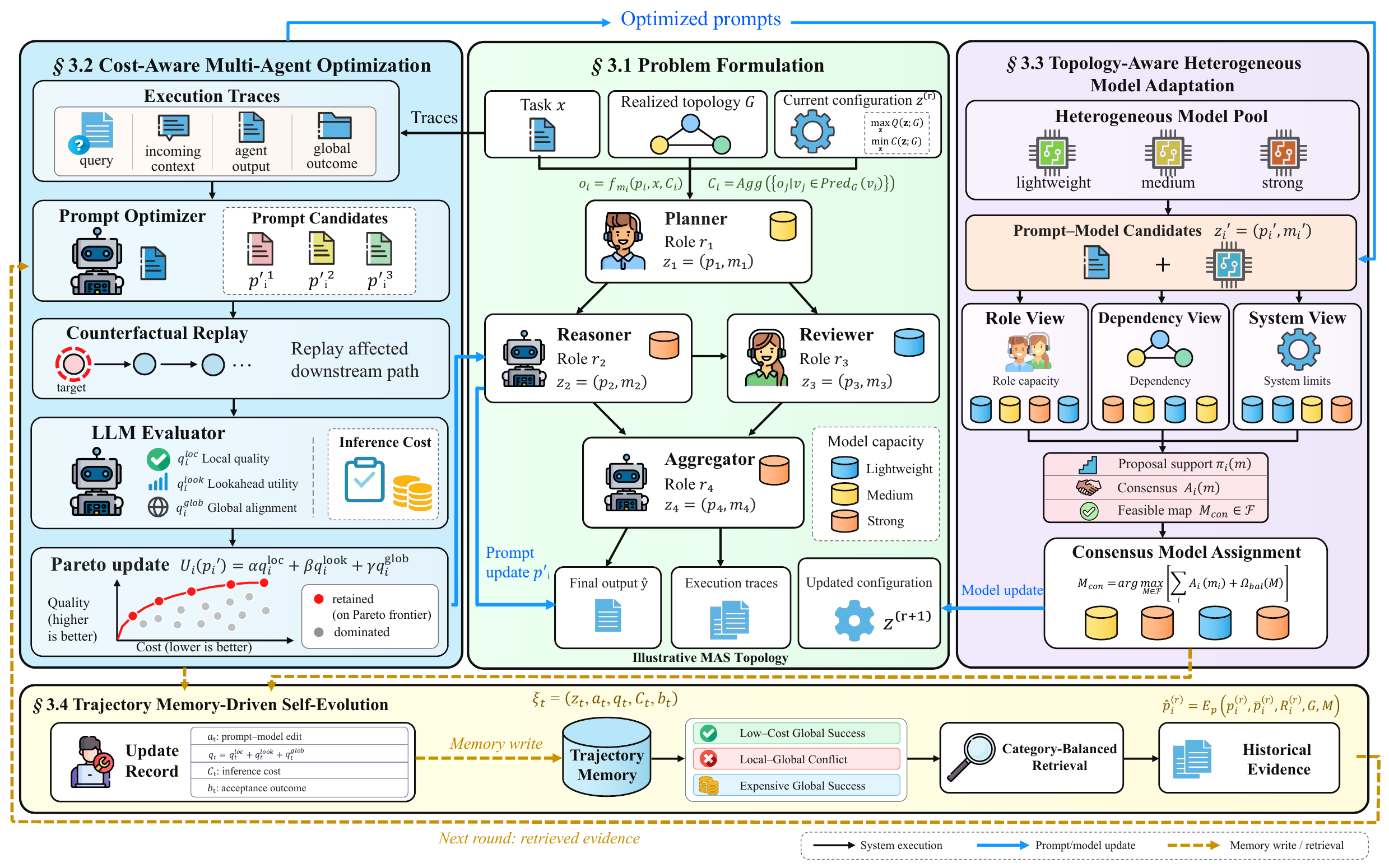}
    \vspace{-13pt}
    \caption{Overview of UniMASO. Cost-aware multi-agent optimization proposes
    and evaluates prompt candidates from execution traces; heterogeneous
    execution runs the current prompt--model configuration on the realized
    topology; topology-aware adaptation reconciles role, dependency, and system
    views to update model assignments; and trajectory memory records
    cross-level feedback and retrieves historical evidence for subsequent
    rounds.}
    \label{fig:unimaso_overview}
\end{figure*}

UniMASO alternates local search over agent configurations with topology-aware
system coordination. As illustrated in Figure~\ref{fig:unimaso_overview}, local
search proposes prompt--model candidates from interaction traces and evaluates
them at multiple granularities, while system-level co-evolution reconciles these
updates under the realized graph. Agents are visited in topological order, and
each candidate is evaluated in context before accepted updates are stored in
trajectory memory for later rounds. This ordering separates local exploration of
prompt--model pairings from their global composition; the complete procedure is
given in Algorithm~\ref{alg:unimaso} in Appendix~\ref{app:algorithm}.

\subsection{Problem Formulation}
\label{sec:problem_formulation}

We model an $N$-agent system as a directed graph $\mathcal G=(\mathcal V,\mathcal E)$.
An edge $(v_j,v_i)\in\mathcal E$ makes the output of $v_j$ available to $v_i$
as context. Each agent $v_i$ has a prompt space $\mathcal P_i$ and is assigned
a role prompt $p_i\in\mathcal P_i$ and an execution model $m_i$ from the
heterogeneous pool $\mathcal M=\{m^{(1)},\ldots,m^{(K)}\}$. We write this
assignment as $z_i=(p_i,m_i)$ and the full system configuration as
$\mathbf z=(z_1,\ldots,z_N)$. Given input $x$, agent $v_i$ produces
\begin{equation}
    o_i=f_{m_i}\!\left(p_i,x,\mathcal C_i\right),
    \qquad
    \mathcal C_i=\operatorname{Agg}\!\left(
        \{o_j\mid v_j\in\operatorname{Pred}_{\mathcal G}(v_i)\}
    \right),
    \label{eq:agent_execution}
\end{equation}
where $\operatorname{Agg}(\cdot)$ is protocol-specific aggregation. The full
execution induces the system prediction $\hat y=\Phi(x;\mathcal G,\mathbf z)$.

Given a validation set $\mathcal D=\{(x_j,y_j)\}_{j=1}^{S}$ and task metric $s$,
let $c(m_i,x_j;\mathbf z,\mathcal G)$ denote the input--output token cost
incurred by agent $v_i$ on example $x_j$. We measure average task quality and
average system cost as
\begin{equation}
    Q(\mathbf z;\mathcal G)=\frac{1}{S}\sum_{j=1}^{S}s\!\left(\Phi(x_j;\mathcal G,\mathbf z),y_j\right),
    \qquad
    C(\mathbf z;\mathcal G)=\frac{1}{S}\sum_{j=1}^{S}\sum_{i=1}^{N}c(m_i,x_j;\mathbf z,\mathcal G).
    \label{eq:quality_cost_objectives}
\end{equation}
The joint objective is to maximize task quality and minimize average token
cost, i.e., $\max_{\mathbf z} Q(\mathbf z;\mathcal G)$ and
$\min_{\mathbf z} C(\mathbf z;\mathcal G)$. Because local prompt or model
changes propagate through $\mathcal G$, they can jointly alter downstream
context, task quality, and cost. UniMASO therefore evaluates prompt--model
changes at the system level.
\subsection{Cost-Aware Multi-Agent Optimization}
\label{sec:agent_level_optimization}

\paragraph{Trace-guided candidate generation.}
An agent's prompt controls how it processes upstream information and what it
passes to successors. For focal agent $v_i$, the optimizer
$\mathcal M_{\mathrm{opt}}$ samples traces containing inputs, contexts, outputs,
and feedback, then proposes candidates from the current configuration. We first
hold the model assignment fixed to measure prompt effects in context. Static
topologies re-execute the focal agent and affected descendants, whereas routing
changes trigger a full interaction; model alternatives are evaluated by the same
procedure in Section~\ref{sec:system_level_coevolution}.

\paragraph{Multi-granularity interaction evaluation.}
Local role fulfillment need not improve the final task. Following MASPO
\citep{wang2026maspo}, we compare candidate and reference executions at three
granularities: local validity, lookahead utility, and global alignment:
\begin{equation}
    U_i(p') =
    \alpha q_i^{\mathrm{loc}}(p')
    + \beta q_i^{\mathrm{look}}(p')
    + \gamma q_i^{\mathrm{glob}}(p'),
    \qquad \alpha+\beta+\gamma=1.
    \label{eq:multilevel_quality}
\end{equation}
Here, $q_i^{\mathrm{loc}}$ measures role fulfillment,
$q_i^{\mathrm{look}}$ measures the effect on successor outputs, and
$q_i^{\mathrm{glob}}$ measures the final response. Each is a normalized
preference over the evaluation batch from $\mathcal M_{\mathrm{eval}}$. Their
weighted sum $U_i$ guides candidate selection; it is distinct from the final
task quality $Q$ defined in Section~\ref{sec:problem_formulation}.

\paragraph{Cost-aware beam search.}
Prompt changes can alter both token use and downstream cost. Building on MASPO's
evolutionary beam search \citep{wang2026maspo}, let $C_i(p')$ be the mean system
cost under candidate $p'$ and $\eta_i(p')=C_i(p')/C_i(p_i)$ its relative cost. We
expand a candidate when $U_i(p')\eta_i(p')^{-\lambda}>u_{\mathrm{ref}}$, where $\lambda\geq0$ is the cost weight. The threshold $u_{\mathrm{ref}}=1/2$ is the neutral point of the pairwise preference score, so candidates that are not preferred to the current prompt after cost adjustment are filtered out.
Eligible candidates and the current prompt form $\mathcal P_i^{+}$, from which
we retain quality--cost nondominated candidates:
\begin{equation}
    \mathcal B_i =
    \operatorname{Select}_{B}\!\left(
        \operatorname{ND}_{U_i\uparrow,\;C_i\downarrow}
        \left(\mathcal P_i^{+}\right)
    \right).
    \label{eq:pareto_beam}
\end{equation}
The operator $\operatorname{ND}$ removes candidates dominated in utility and
cost, while $\operatorname{Select}_{B}$ keeps at most $B$ candidates. Cost
calibration controls expansion and nondominated selection preserves distinct
quality--cost trade-offs. Retained candidates are re-evaluated after peer
updates, and the best local state is passed to system coordination.

\subsection{Topology-Aware Heterogeneous Model Adaptation}
\label{sec:system_level_coevolution}

A fixed assignment limits the capabilities available at each node. We therefore
extend the local candidates to prompt--model pairs
$z_i'=(p_i',m_i')$, $m_i'\in\mathcal M$. Model capability and cost condition
candidate generation, and every pair is executed and scored with the same
multi-granularity utility and cost rule. The best local update is
$\bar z_i=(\bar p_i,\bar m_i)$, which is then passed to system coordination.

\paragraph{Topology-conditioned multi-view proposals.}
Local updates are evaluated with the remaining configuration fixed, so their
combination may change inter-agent behavior. System-level evolution resolves
this mismatch: each optimizable logical agent forms a focal view conditioned on
roles, dependencies, execution order, local search results, and model
capability--cost information. It proposes a prompt $\hat p_j$ and complete
assignment $\hat{\mathbf M}^{(j)}$; shared logical roles share variables, while
deterministic routing and aggregation components remain fixed.

\paragraph{Proposal-entropy consensus.}
Focal views may disagree on a node's model. We measure their empirical support
and weight it by proposal concentration:
\begin{equation}
    \pi_i(m)=\frac{1}{J}\sum_{j=1}^{J}
    \mathbb I[\hat m_i^{(j)}=m],
    \qquad
    A_i(m)=\pi_i(m)(1-\bar H_i).
    \label{eq:model_support}
\end{equation}
Here, $J$ is the number of focal views and $\bar H_i$ is the Shannon entropy of
$\pi_i$ normalized by $\log|\mathcal M_i|$. We set $\bar H_i=0$ when only one
model is feasible. Concentrated proposals therefore receive greater weight in
the configuration-level decision, which is solved subject to feasibility and
capability--cost balance:
\begin{equation}
    \mathbf M_{\mathrm{con}} =
    \arg\max_{\mathbf M\in\mathcal F}
    \left[
        \sum_{i=1}^{N} A_i(m_i)
        + \Omega_{\mathrm{bal}}(\mathbf M)
    \right].
    \label{eq:consensus_assignment}
\end{equation}
The feasible set $\mathcal F$ enforces model availability, role-sharing, and
routing constraints. We define $\Omega_{\mathrm{bal}}(\mathbf M)=
g_{\theta}(|\mathcal M|)\,B(\mathbf M;\mathcal M)$, where $\mathcal M$ is the
model pool, $B$ scores capability--cost balance, and $g_{\theta}$ sets its
pool-size-dependent weight. This assignment guides evaluation and subsequent
search.

\subsection{Trajectory Memory-Driven Self-Evolution}
\label{sec:trajectory_memory}

Local search and system coordination produce trajectories linking configuration
changes to local, downstream, and global outcomes. We store these records as
retrievable experience for later evolution.

\paragraph{Cross-level trajectory encoding.}
For optimization transition $t$, we record the preceding configuration, the
proposed modification, and its evaluation:
\begin{equation}
    \xi_t=(\mathbf z_t,a_t,\mathbf q_t,C_t,b_t),
    \qquad
    \mathbf q_t=
    (q_t^{\mathrm{loc}},q_t^{\mathrm{look}},q_t^{\mathrm{glob}}).
    \label{eq:trajectory_encoding}
\end{equation}
Here, $\mathbf z_t$ is the preceding configuration, $a_t$ the modification,
$\mathbf q_t$ the multi-granularity feedback, $C_t$ the execution cost, and
$b_t$ the acceptance outcome. We group records into low-cost global successes,
local successes with global failures, and high-cost global successes, providing
examples of efficient changes, coordination conflicts, and effective but costly
alternatives.

\paragraph{Category-balanced retrieval.}
To prevent frequent categories from dominating nearest-neighbor retrieval, we
retrieve within each category. The normalized embedding of the current prompt
selects up to $K_{\mathrm{mem}}$ trajectories per category by similarity. Each
record contains the configuration, modification, and evaluation feedback.

\paragraph{Experience-guided evolution.}
At round $r$, retrieved experience is combined with the current and locally
optimized prompts:
\begin{equation}
    \hat p_i^{(r)} =
    \mathcal E_p\!\left(
        p_i^{(r)},\bar p_i^{(r)},
        \mathcal R_i^{(r)},\mathcal G,\mathcal M
    \right).
    \label{eq:memory_guided_prompt_evolution}
\end{equation}
The evolution operator $\mathcal E_p$ interprets these records under current
dependencies and available models. Low-cost successes suggest reusable changes,
local--global conflicts expose downstream effects, and high-cost successes
provide targets for cost reduction. The updated configuration is executed and
its trajectory is written back to memory.

\section{Experiments}
\label{sec:experiments}

\begin{table*}[t]
\centering
\scriptsize
\setlength{\tabcolsep}{3.8pt}
\caption{Task quality and deployment cost under the centralized topology.
Prompt-only baselines use homogeneous models; UniMASO jointly optimizes prompts
and model assignments.}
\label{tab:main_results}
\resizebox{0.9\textwidth}{!}{%
\begin{tabular}{>{\raggedright\arraybackslash}m{2.35cm}@{\hskip 1pt}l@{\hskip 4pt}c@{\hskip 3pt}c@{\hskip 6pt}c@{\hskip 3pt}c@{\hskip 6pt}c@{\hskip 3pt}c@{\hskip 6pt}c@{\hskip 3pt}c@{\hskip 6pt}c@{\hskip 3pt}c@{\hskip 6pt}c@{\hskip 3pt}c}
\toprule
\textbf{Model} & \textbf{Method}
& \multicolumn{2}{c}{\textbf{AGIEval}} & \multicolumn{2}{c}{\textbf{GPQA}} & \multicolumn{2}{c}{\textbf{BFCL}} & \multicolumn{2}{c}{\textbf{LiveCodeBench}} & \multicolumn{2}{c}{\textbf{HotpotQA}} & \multicolumn{2}{c}{\textbf{Avg.}} \\
& & \textbf{Acc.} & \textbf{Cost} & \textbf{Acc.} & \textbf{Cost} & \textbf{Acc.} & \textbf{Cost} & \textbf{pass@1} & \textbf{Cost} & \textbf{EM} & \textbf{Cost} & \textbf{Result} & \textbf{Cost} \\
\midrule

\multirow{5}{*}{Qwen3.5-4B}
& Vanilla MAS & 62.11 & 7.3 & 67.17 & 5.3 & 88.40 & 4.8 & 18.32 & 3.8 & 58.00 & 5.8 & 58.80 & 5.38 \\
& \hspace{0.4em}+TPE & 59.77 & 8.2 & 66.67 & 6.1 & 88.80 & 5.4 & 19.08 & 4.2 & 58.20 & 6.5 & 58.50 & 6.08 \\
& \hspace{0.4em}+SPO & 63.28 & 8.3 & 69.70 & 5.9 & 88.60 & 5.5 & 20.61 & 4.3 & 58.00 & 6.6 & 60.03 & 6.12 \\
& \hspace{0.4em}+MASPOB & 63.67 & 8.2 & 68.69 & 5.8 & 89.40 & 5.3 & 19.85 & 4.3 & 58.60 & 6.5 & 60.04 & 6.02 \\
& \hspace{0.4em}+MASPO & 63.28 & 8.2 & 69.19 & 5.9 & 89.60 & 5.5 & 21.37 & 4.4 & 58.40 & 6.4 & 60.36 & 6.08 \\
\midrule

\multirow{5}{*}{Qwen3.5-9B}
& Vanilla MAS & 71.09 & 14.3 & 71.72 & 12.2 & 90.40 & 9.1 & 22.14 & 8.1 & 56.80 & 10.5 & 62.43 & 10.84 \\
& \hspace{0.4em}+TPE & 71.88 & 15.8 & 72.22 & 13.6 & 87.80 & 10.2 & 21.37 & 9.1 & 57.20 & 12.0 & 62.09 & 12.14 \\
& \hspace{0.4em}+SPO & 71.48 & 16.3 & 73.23 & 13.7 & 90.80 & 10.2 & 25.19 & 8.9 & 57.00 & 11.6 & 63.54 & 12.14 \\
& \hspace{0.4em}+MASPOB & 72.66 & 16.4 & 72.73 & 13.7 & 91.00 & 10.0 & 24.43 & 9.3 & 57.60 & 11.6 & 63.68 & 12.20 \\
& \hspace{0.4em}+MASPO & 72.27 & 16.2 & 73.74 & 13.6 & 91.40 & 10.1 & 25.19 & 9.0 & 57.40 & 11.8 & 64.00 & 12.18 \\
\midrule

\multirow{5}{*}{Qwen3.5-27B}
& Vanilla MAS & 75.78 & 56.2 & 77.27 & 41.0 & 90.40 & 38.7 & 29.77 & 26.4 & 60.80 & 48.5 & 66.80 & 42.18 \\
& \hspace{0.4em}+TPE & 76.17 & 64.2 & 78.28 & 46.2 & 91.00 & 44.5 & 30.53 & 29.3 & 63.80 & 54.1 & 67.96 & 47.66 \\
& \hspace{0.4em}+SPO & 73.05 & 64.1 & 77.78 & 46.1 & 90.60 & 43.6 & 29.01 & 30.0 & 61.20 & 54.4 & 66.33 & 47.64 \\
& \hspace{0.4em}+MASPOB & 76.56 & 62.1 & 79.29 & 46.9 & 91.40 & 43.6 & 32.06 & 29.7 & 61.40 & 55.6 & 68.14 & 47.58 \\
& \hspace{0.4em}+MASPO & 77.34 & 63.5 & 78.79 & 45.7 & 91.60 & 43.4 & 31.30 & 30.1 & 61.80 & 54.5 & 68.17 & 47.44 \\
\midrule

\multirow{5}{*}{DeepSeek-V4-Flash}
& Vanilla MAS & 72.27 & 57.3 & 70.20 & 24.8 & 92.20 & 73.1 & 38.17 & 39.6 & 59.20 & 100.2 & 66.41 & 59.00 \\
& \hspace{0.4em}+TPE & 72.66 & 64.2 & 67.68 & 27.4 & 92.60 & 81.5 & 38.93 & 44.6 & 59.60 & 112.7 & 66.30 & 66.08 \\
& \hspace{0.4em}+SPO & 73.05 & 64.1 & 70.71 & 27.7 & 92.40 & 81.5 & 40.46 & 43.8 & 62.00 & 114.7 & 67.72 & 66.36 \\
& \hspace{0.4em}+MASPOB & 73.83 & 64.0 & 71.72 & 28.0 & 92.80 & 82.8 & 39.69 & 45.5 & 60.00 & 114.5 & 67.61 & 66.96 \\
& \hspace{0.4em}+MASPO & 73.44 & 63.3 & 72.22 & 28.1 & 93.20 & 83.4 & 41.22 & 44.0 & 59.80 & 113.1 & 67.97 & 66.38 \\
\midrule

\multirow{5}{*}{DeepSeek-V4-Pro}
& Vanilla MAS & 76.17 & 217.3 & 78.28 & 103.4 & 92.00 & 195.1 & 32.82 & 112.0 & 61.60 & 295.0 & 68.17 & 184.55 \\
& \hspace{0.4em}+TPE & 76.95 & 249.3 & 78.79 & 118.6 & 92.20 & 219.9 & 30.53 & 128.7 & 62.00 & 330.7 & 68.09 & 209.44 \\
& \hspace{0.4em}+SPO & 76.56 & 239.2 & 79.80 & 116.6 & 94.80 & 219.6 & 33.59 & 125.2 & 61.80 & 335.8 & 69.31 & 207.28 \\
& \hspace{0.4em}+MASPOB & 77.34 & 246.9 & 79.29 & 117.2 & 92.80 & 219.2 & 35.88 & 125.8 & 62.60 & 338.4 & 69.58 & 209.50 \\
& \hspace{0.4em}+MASPO & 78.12 & 249.6 & 80.30 & 115.2 & 92.60 & 216.6 & 35.11 & 124.0 & 62.80 & 336.0 & 69.79 & 208.28 \\
\midrule
\multirow{4}{*}{DeepSeek-V4-Pro}
& Single Agent & 74.61 & 36.5 & 76.77 & 23.0 & 92.80 & 25.9 & 36.64 & 16.9 & 52.60 & 55.2 & 66.68 & 31.50 \\
& \hspace{0.4em}+CoT & 75.39 & 42.3 & 75.76 & 26.5 & 85.00 & 26.6 & 38.93 & 21.9 & 61.00 & 69.8 & 67.22 & 37.41 \\
& \hspace{0.4em}+TPE & 71.09 & 37.2 & 73.74 & 16.1 & 92.60 & 26.5 & 36.64 & 16.3 & 61.20 & 54.2 & 67.05 & 30.04 \\
& \hspace{0.4em}+SPO & 75.39 & 39.9 & 73.74 & 16.9 & 93.00 & 26.4 & 38.93 & 16.0 & 60.20 & 54.1 & 68.25 & 30.65 \\
\midrule
Mixed Models & UniMASO & 82.42 & 56.9 & 78.28 & 33.0 & 92.40 & 12.2 & 35.11 & 13.3 & 63.80 & 39.3 & 70.40 & 30.92 \\
\bottomrule
\end{tabular}
}
\end{table*}

\subsection{Experimental Setup}
\label{sec:experimental_setup}

\paragraph{Benchmarks, Models, and Topologies.}
We evaluate UniMASO on AGIEval-MATH at difficulty level 5
\citep{zhong-etal-2024-agieval}, GPQA-Diamond \citep{rein2024gpqa},
BFCL-V3 \citep{patil2025bfcl,yan2025bfcl}, LiveCodeBench v6
\citep{jain2025livecodebench}, and HotpotQA
\citep{yang-etal-2018-hotpotqa}. LiveCodeBench uses problems released
between February 1 and May 1, 2025. We test Independent, Centralized,
Sequential, Decentralized, and Dynamic topologies to vary information flow
and downstream dependencies. The model
pool comprises Qwen3.5-4B, Qwen3.5-9B, Qwen3.5-27B \citep{qwen35},
DeepSeek-V4-Flash, and DeepSeek-V4-Pro \citep{deepseekv4_modelcard}.
Dataset details and topology configurations appear in
Appendices~\ref{app:benchmarks} and~\ref{app:topologies}, respectively.

\paragraph{Baselines.}
TPE (Tree-structured Parzen Estimator), used in MIPRO~\citep{opsahlong2024optimizing}
and MASS~\citep{zhou2026multiagent}, and SPO (Self-Supervised Prompt
Optimization)~\citep{xiang-etal-2025-spo} are prompt optimizers. We compare
UniMASO with DeepSeek-V4-Pro single-agent baselines (direct, CoT, +TPE, +SPO)
and MAS baselines (unoptimized Centralized MAS, TPE, SPO,
MASPOB~\citep{hong2026maspob}, and MASPO~\citep{wang2026maspo}). MAS prompt-only
baselines optimize role prompts while keeping topology and model assignments
fixed. Across topologies, five homogeneous configurations are compared with
UniMASO's heterogeneous assignment under identical evaluation and
deployment-cost accounting.

\paragraph{Implementation Details.}
Unless otherwise stated, UniMASO runs three optimization rounds with per-round
search depths of $3$--$2$--$2$. Task agents use temperature $0.0$, and optimizer
and evaluator settings are fixed across methods. Following
MASPO~\citep{wang2026maspo}, we maintain an optimization pool of
$|\mathcal D_{\mathrm{opt}}|=50$ per benchmark and randomly sample 10 instances
per iteration. Memory retrieval selects up to $K_{\mathrm{mem}}=2$ trajectories
per category, and the cost weight is set to $\lambda=0.2$. The selected
configuration is evaluated on each benchmark's full evaluation file.
Further implementation details and prompt templates appear in
Appendices~\ref{app:implementation_details} and~\ref{app:prompts}, respectively.

\paragraph{Metrics.}
We report accuracy for AGIEval-MATH, GPQA-Diamond, and BFCL-V3, Pass@1 for
LiveCodeBench, and exact match for HotpotQA. The Avg. columns report means of
the five benchmark scores and their corresponding total deployment costs,
respectively. We define task-agent deployment cost as
$\mathrm{Cost}(\mathbf z)=\sum_i(c_{m_i}^{\mathrm{in}}T_i^{\mathrm{in}}+
c_{m_i}^{\mathrm{out}}T_i^{\mathrm{out}})$, where $T_i^{\mathrm{in}}$ and
$T_i^{\mathrm{out}}$ are the input and output token counts of agent $i$, and
$c_{m_i}^{\mathrm{in}}$ and $c_{m_i}^{\mathrm{out}}$ are fixed relative
model-cost coefficients derived from the Qwen AI model pricing reference.
We report the cost of executing the selected configuration in millions of
weighted-token units (M). For each benchmark, the reported cost is summed
over its full evaluation set. Further details appear in
Appendix~\ref{app:metrics}.

\subsection{Main Results}
\label{sec:main_results}

\begin{minipage}{\linewidth}
\paragraph{UniMASO achieves the strongest centralized quality--cost trade-off.}
Table~\ref{tab:main_results} compares UniMASO with prompt-only baselines on the
same centralized topology. These baselines optimize role prompts while keeping
model assignments homogeneous. Within each model group, the best prompt-only
variant improves on its unoptimized counterpart but incurs higher deployment
cost. DeepSeek-V4-Pro + MASPO is the strongest prompt-only configuration,
scoring 69.79 at a cost of 208.28M. UniMASO raises the average score to 70.40
and lowers cost to 30.92M, a gain of 0.61 points with an 85.2\% reduction in
token-weighted deployment cost. It achieves this operating point by jointly
adapting prompts and model assignments to match model capacity to agent roles.
UniMASO leads on AGIEval-MATH (82.42 vs. 78.12) and HotpotQA (63.80 vs. 62.80),
while homogeneous configurations score higher on GPQA-Diamond, BFCL-V3, and
LiveCodeBench.
\end{minipage}

\begin{figure}[t]
    \setlength{\abovecaptionskip}{1pt}
    \setlength{\belowcaptionskip}{0pt}
    \begin{minipage}[t]{0.49\textwidth}
        \centering
        \includegraphics[width=\linewidth]{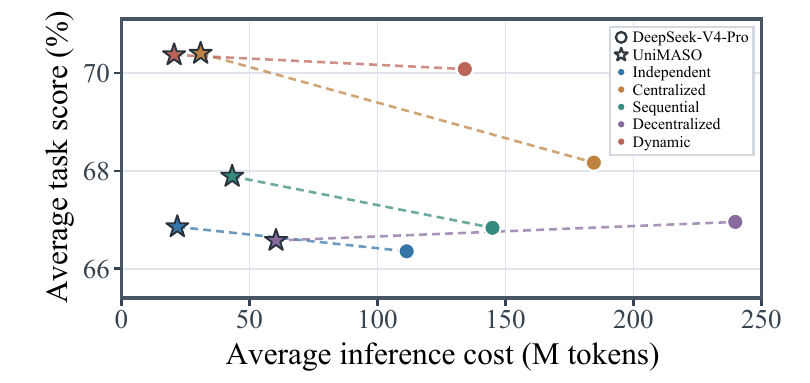}
        \vspace{-7pt}
        \caption{Quality--cost trade-off by topology.}
        \label{fig:topology_tradeoff}
    \end{minipage}\hfill
    \begin{minipage}[t]{0.49\textwidth}
        \centering
        \includegraphics[width=\linewidth]{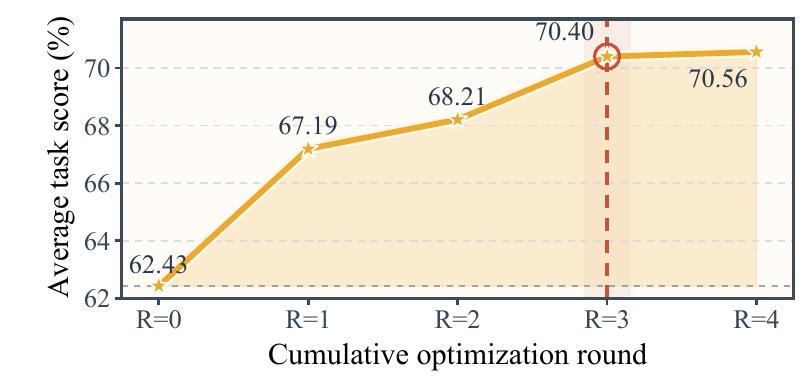}
        \vspace{-7pt}
        \caption{Mean score by optimization round.}
        \label{fig:roundwise_gain}
    \end{minipage}
\end{figure}

\paragraph{UniMASO's quality--cost advantage transfers across execution topologies.}
Figure~\ref{fig:topology_tradeoff} plots the paired quality--cost points for
DeepSeek-V4-Pro and UniMASO across Independent, Centralized, Sequential,
Decentralized, and Dynamic execution. UniMASO moves the comparison toward
lower deployment cost in all five topologies and improves the mean score in
four. Decentralized execution is the only quality decrease, trading 66.96 for
66.58 while reducing cost from 239.80M to 60.37M. The Dynamic topology gives a
representative example of the same pattern: UniMASO reaches 70.37 at 20.56M,
compared with 70.08 at 134.13M for the homogeneous reference. Full per-task
scores and costs appear in Table~\ref{tab:topology_results}
(Appendix~\ref{app:topologies}).

\section{Analysis}
\label{sec:analysis}

We analyze the contributions of UniMASO's components, the sensitivity of its
optimization settings, and the behavior of the resulting MAS. Unless stated
otherwise, comparisons use centralized execution and the protocol in
Section~\ref{sec:experimental_setup}. Table~\ref{tab:ablation_results} reports
component ablations and parameter sensitivity.

\subsection{Ablation Study of UniMASO}
\label{sec:ablation_study}

\paragraph{Effect of Joint Prompt--Model Optimization.}
We test whether role prompts and model assignments should be optimized
together rather than in consecutive stages. The control first optimizes
prompts and then assigns models, whereas UniMASO evaluates prompt--model
pairs during joint search under the same centralized topology
(Table~\ref{tab:ablation_results}). The staged control lowers the average
score from 70.40 to 66.39 and raises deployment cost from 30.92M to
64.15M. The loss spans all five benchmarks; on GPQA-Diamond, accuracy
falls from 78.28 to 71.72 while cost rises from 33.02M to 70.24M. A prompt
chosen before model reassignment may fit the original executor better than
the model ultimately assigned to that role. Jointly evaluating both
decisions is therefore consistent with the observed improvement in
quality and cost.

\paragraph{Effect of Topology-Aware Co-Evolution.}
We examine whether locally selected updates remain effective after
interacting agents adopt their new prompts and models. The ablation
removes system-level co-evolution while retaining local candidate
search on the centralized graph (Table~\ref{tab:ablation_results}).
Without cross-agent coordination, the average score falls from 70.40
to 68.49, whereas deployment cost changes only from 30.92M to 31.20M.
GPQA-Diamond shows the largest score drop (4.54 points), followed by
AGIEval-MATH (2.34 points) and LiveCodeBench (2.29 points); HotpotQA
improves slightly (0.20 points). The near-constant cost suggests that the main
difference is how updates interact across dependent roles rather
than a systematic shift toward more expensive execution. Coordinating
local proposals is useful when a change at one agent affects its
successors.

\begin{table*}[t]
\centering
\scriptsize
\setlength{\tabcolsep}{2.8pt}
\newsavebox{\unimasoablationwidth}
\newif\ifunimasoablationmeasure
\makeatletter
\newcommand{\unimasoablationrowskip}{\dimexpr\ht\@arstrutbox+\dp\@arstrutbox\relax}
\makeatother
\newcommand{\unimasoablationsection}[1]{%
  \multicolumn{13}{@{}l@{}}{\textit{#1}} \\
  \ifunimasoablationmeasure\else
  \multicolumn{13}{@{}c@{}}{%
    \raisebox{\dimexpr\unimasoablationrowskip*4/5\relax}[0pt][0pt]{%
      \makebox[\wd\unimasoablationwidth][l]{%
        \cleaders\hbox{\hskip.175em\rule{.6em}{.4pt}\hskip.175em}\hfill}}}%
  \\[-\unimasoablationrowskip]
  \fi
}
\caption{UniMASO ablations and parameter sensitivity under centralized execution.}
\label{tab:ablation_results}
\newcommand{\unimasoablationtable}{%
\begin{tabular}{>{\raggedright\arraybackslash}m{3.60cm}@{\hskip 1pt}c@{\hskip 2pt}c@{\hskip 4pt}@{\hskip 1pt}c@{\hskip 2pt}c@{\hskip 4pt}@{\hskip 1pt}c@{\hskip 2pt}c@{\hskip 4pt}@{\hskip 1pt}c@{\hskip 2pt}c@{\hskip 4pt}@{\hskip 1pt}c@{\hskip 2pt}c@{\hskip 4pt}@{\hskip 1pt}c@{\hskip 2pt}c}
\toprule
\textbf{Method}
& \multicolumn{2}{c}{\textbf{AGIEval}}
& \multicolumn{2}{c}{\textbf{GPQA}}
& \multicolumn{2}{c}{\textbf{BFCL}}
& \multicolumn{2}{c}{\textbf{LiveCodeBench}}
& \multicolumn{2}{c}{\textbf{HotpotQA}}
& \multicolumn{2}{c}{\textbf{Avg.}} \\
& \textbf{Acc.} & \textbf{Cost}
& \textbf{Acc.} & \textbf{Cost}
& \textbf{Acc.} & \textbf{Cost}
& \textbf{pass@1} & \textbf{Cost}
& \textbf{EM} & \textbf{Cost}
& \textbf{Result} & \textbf{Cost} \\
\midrule

\unimasoablationsection{Reference: Proposed Framework}

UniMASO (Full)
& 82.42 & 56.86
& 78.28 & 33.02
& 92.40 & 12.21
& 35.11 & 13.27
& 63.80 & 39.26
& 70.40 & 30.92 \\

\midrule

\unimasoablationsection{(I) Component Ablations}

w/o Memory
& 80.08 & 51.35
& 76.77 & 31.58
& 91.60 & 13.18
& 29.01 & 13.35
& 61.80 & 23.04
& 67.85 & 26.50 \\

w/o Co-Evolution
& 80.08 & 53.31
& 73.74 & 37.57
& 91.80 & 13.12
& 32.82 & 13.04
& 64.00 & 38.96
& 68.49 & 31.20 \\

w/o Joint Optimization
& 78.52 & 121.38
& 71.72 & 70.24
& 89.20 & 18.07
& 30.53 & 24.89
& 62.00 & 86.17
& 66.39 & 64.15 \\

w/o Cost-Aware Pareto Search
& 82.03 & 110.81
& 76.26 & 73.28
& 92.60 & 24.53
& 33.59 & 26.22
& 62.60 & 77.16
& 69.42 & 62.40 \\

\midrule

\unimasoablationsection{(II) Optimization Depth per Round (R1--R2--R3; default: 3--2--2)}

Depth = 3--3--3
& 82.03 & 109.49
& 77.78 & 79.97
& 92.20 & 34.05
& 35.11 & 24.56
& 63.60 & 92.93
& 70.14 & 68.20 \\

Depth = 3--2--1
& 78.12 & 52.37
& 74.24 & 31.94
& 91.40 & 13.07
& 28.24 & 9.94
& 62.40 & 20.68
& 66.88 & 25.60 \\

\midrule

\unimasoablationsection{(III) Sensitivity to Retrieved Memory Size ($K_{\mathrm{mem}}$, default: 2)}

$K_{\mathrm{mem}}=1$
& 80.86 & 44.65
& 77.78 & 22.92
& 90.60 & 9.99
& 29.01 & 12.26
& 64.00 & 50.68
& 68.45 & 28.10 \\

$K_{\mathrm{mem}}=3$
& 80.86 & 90.15
& 77.78 & 69.55
& 92.00 & 24.96
& 35.88 & 22.51
& 62.80 & 76.83
& 69.86 & 56.80 \\

$K_{\mathrm{mem}}=4$
& 79.30 & 138.32
& 75.76 & 69.87
& 91.00 & 18.20
& 28.24 & 28.07
& 61.20 & 67.04
& 67.10 & 64.30 \\

\midrule

\unimasoablationsection{(IV) Sensitivity to Cost Weight ($\lambda$, default: 0.2)}

$\lambda=0$
& 81.64 & 110.24
& 77.78 & 73.18
& 92.00 & 18.65
& 34.35 & 21.47
& 63.60 & 80.36
& 69.87 & 60.78 \\

$\lambda=0.1$
& 81.25 & 69.82
& 77.78 & 43.26
& 91.80 & 15.72
& 35.88 & 20.64
& 63.40 & 76.13
& 70.02 & 45.11 \\

$\lambda=0.5$
& 82.03 & 81.47
& 77.27 & 49.38
& 92.20 & 12.16
& 34.35 & 23.94
& 63.60 & 79.12
& 69.89 & 49.21 \\

$\lambda=1.0$
& 80.86 & 74.25
& 77.78 & 45.17
& 92.00 & 19.18
& 35.11 & 21.73
& 63.40 & 73.84
& 69.83 & 46.83 \\
\bottomrule
\end{tabular}%
}
\unimasoablationmeasuretrue
\sbox{\unimasoablationwidth}{\unimasoablationtable}
\unimasoablationmeasurefalse
\resizebox{0.9\textwidth}{!}{\unimasoablationtable}
\end{table*}

\paragraph{Effect of Cost-Aware Pareto Search.}
We assess whether candidate selection needs to account for execution
cost as well as task utility. UniMASO uses relative cost when expanding
candidates and retains nondominated quality--cost states; the ablation
removes this cost-aware Pareto search
(Table~\ref{tab:ablation_results}). The average score decreases from
70.40 to 69.42, while deployment cost increases from 30.92M to
62.40M. On BFCL-V3, the ablation raises accuracy slightly from 92.40
to 92.60 but almost doubles cost from 12.21M to 24.53M. AGIEval-MATH
shows a similar cost increase, from 56.86M to 110.81M, without an
accuracy gain. These comparisons show why task quality alone does not
identify an efficient configuration. Cost-aware expansion and
retention improve the average operating point across benchmarks.

\begin{wraptable}[8]{r}{0.36\textwidth}
    \vspace{-0.7\baselineskip}
    \centering
    \caption{\footnotesize Cross-task memory transfer.}
    \label{tab:memory_transfer}
    \vspace{2pt}
    \fontsize{6.5}{7.4}\selectfont
    \setlength{\tabcolsep}{2pt}
    \renewcommand{\arraystretch}{0.96}
    \begin{tabular}{@{}m{1.72cm}m{0.82cm}cc@{}}
        \toprule
        \textbf{Target} & \textbf{Memory} & \textbf{Acc.} & \textbf{Cost} \\
        \midrule
        \multirow{3}{*}{AGIEval-MATH}
        & None & 80.08 & 51.35 \\
        & Own & 82.42 & 56.86 \\
        & Other & 81.64 & 52.93 \\
        \midrule
        \multirow{3}{*}{GPQA-Diamond}
        & None & 76.77 & 31.58 \\
        & Own & 78.28 & 33.02 \\
        & Other & 77.78 & 29.43 \\
        \bottomrule
    \end{tabular}
\end{wraptable}

\paragraph{Effect of Trajectory Memory.}
We test whether stored trajectories improve optimization and
transfer across related tasks. Removing memory lowers the average
score from 70.40 to 67.85 and cost from 30.92M to 26.50M
(Table~\ref{tab:ablation_results}); LiveCodeBench drops most, from
35.11 to 29.01. \mbox{Table~\ref{tab:memory_transfer}} compares no memory,
own-task records, and cross-task records on AGIEval-MATH and
GPQA-Diamond. Cross-task records raise scores from 80.08 to
81.64 and from 76.77 to 77.78, respectively, retaining roughly
two-thirds of the gains from own-task memory. Cross-task costs of
52.93M and 29.43M are below own-task costs of 56.86M and 33.02M.
Memory transfers across tasks, while own-task records remain most accurate.

\subsection{Optimization Dynamics and Quality--Cost Trade-offs}
\label{sec:optimization_dynamics}

\paragraph{Effect of Optimization Rounds.}
We vary the number of complete optimization rounds to determine when
additional adaptation yields diminishing returns. Figure~\ref{fig:roundwise_gain}
tracks the average task score through four rounds; the selected
configuration uses three. The score reaches 67.19 after the first
round, 68.21 after the second, and 70.40 after the third. A fourth
round adds only 0.16 points, reaching 70.56. Thus, the first three
rounds capture 98.0\% of the four-round gain, making round three a
practical stopping point for these tasks.

\paragraph{Effect of Per-Round Search Budget.}
We separately vary the number of candidate-search iterations
allocated to each of the three rounds. The default 3--2--2
schedule is compared with 3--3--3 and 3--2--1 while trajectory
memory remains enabled (Table~\ref{tab:ablation_results}).
With 3--3--3, the selected configuration scores 70.14 at
68.20M, versus 70.40 at 30.92M for the default: deployment
cost more than doubles without a score gain. The 3--2--1
schedule lowers cost to 25.60M but also lowers the score to
66.88. Since the number of rounds is fixed, these differences reflect
the per-round budget. Among these schedules, 3--2--2 yields the best
quality--cost balance.

\paragraph{Effect of Retrieved Memory Size.}
We vary how many trajectories are retrieved from each memory
category to examine the amount of history used during evolution.
The default is $K_{\mathrm{mem}}=2$; we also evaluate one, three,
and four records per category under centralized execution
(Table~\ref{tab:ablation_results}). With one record, the average
score falls from 70.40 to 68.45, while cost declines from
30.92M to 28.10M. Three records yield 69.86 at 56.80M,
and four yield 67.10 at 64.30M. Although three records improve
LiveCodeBench from 35.11 to 35.88, neither larger setting
improves the overall score. Two records give the strongest average
without the higher deployment costs of larger retrieval contexts.

\paragraph{Effect of Cost Weight.}
We vary the search weight $\lambda$ to determine how strongly
relative execution cost should affect candidate expansion.
The default $\lambda=0.2$ is compared with 0, 0.1, 0.5,
and 1.0 in Table~\ref{tab:ablation_results}. Removing the
cost term yields 69.87 at 60.78M, versus 70.40 at 30.92M
for the default. Weights 0.1, 0.5, and 1.0 yield 70.02
at 45.11M, 69.89 at 49.21M, and 69.83 at 46.83M,
respectively. The default thus attains both the highest
score and the lowest deployment cost in this sweep.
The nonmonotonic response reflects which prompt--model candidates survive
under each weight.

\subsection{Coordination Efficiency and Model Behavior}
\label{sec:behavioral_analysis}

\newcommand{\coordblank}{\textemdash}

\begin{wraptable}[11]{r}{0.45\textwidth}
    \vspace{-0.7\baselineskip}
    \centering
    \scriptsize
    \setlength{\tabcolsep}{2.0pt}
    \renewcommand{\arraystretch}{1.05}
    \setlength{\abovecaptionskip}{0pt}
    \setlength{\belowcaptionskip}{2pt}
    \caption{Coordination and efficiency.}
    \label{tab:coordination_efficiency}
    \resizebox{\linewidth}{!}{%
    \begin{tabular}{@{}>{\raggedright\arraybackslash}p{2.15cm} c@{\hskip 3pt}c@{\hskip 3pt}c@{}}
        \toprule
        \textbf{Method}
        & \multicolumn{2}{c}{\textbf{Coordination}}
        & \multicolumn{1}{c}{\textbf{Efficiency}} \\
        \cmidrule(lr){2-3}\cmidrule(lr){4-4}
        & \textbf{Comm. $\uparrow$}
        & \textbf{Plan. $\uparrow$}
        & \textbf{Cost-of-Pass $\downarrow$} \\
        \midrule

        Centralized MAS & 68.45 & 64.39 & 1.154 \\
        \hspace{0.8em}+TPE & 71.81 & 66.86 & 1.357 \\
        \hspace{0.8em}+SPO & 69.90 & 64.15 & 1.271 \\
        \hspace{0.8em}+MASPOB & 71.54 & 67.08 & 1.245 \\
        \hspace{0.8em}+MASPO & 72.84 & \textbf{68.41} & 1.241 \\
        \midrule
        \textbf{UniMASO}
        & \textbf{73.76} & 67.68 & \textbf{0.184} \\
    \bottomrule
    \end{tabular}
    }
\end{wraptable}

\paragraph{Multi-Agent Coordination and Efficiency.}
We assess \mbox{UniMASO's} coordination and cost per correct answer.
Communication and Planning Scores are reported on a percentage scale using
MultiAgentBench~\citep{zhu-etal-2025-multiagentbench}. Cost-of-Pass takes the
unweighted mean of deployment cost per correct answer across five tasks, in
millions of weighted-token units~\citep{erol2026costofpass}
(Table~\ref{tab:coordination_efficiency}). Compared with centralized +MASPO,
\mbox{UniMASO} raises Communication Score from 72.84 to 73.76 and reduces
Cost-of-Pass from 1.241 to 0.184 (85.2\%), while maintaining a comparable
Planning Score (67.68 versus 68.41). UniMASO thus improves agent communication
and substantially lowers the cost per correct answer.

\begin{figure}[H]
    \setlength{\abovecaptionskip}{1pt}
    \setlength{\belowcaptionskip}{0pt}
    \makebox[\textwidth][c]{%
    \begin{minipage}[t]{0.49\textwidth}
        \centering
        \includegraphics[width=\linewidth]{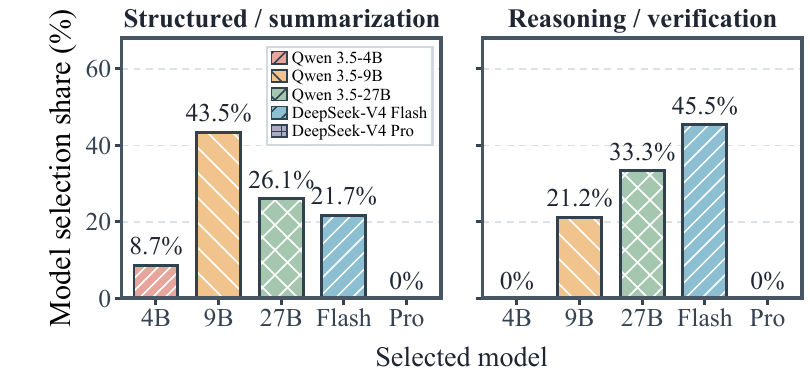}
        \vspace{-7pt}
        {\scriptsize
        \begin{minipage}{0.97\linewidth}
        \caption{Model selection by agent role.}
        \label{fig:role_allocation}
        \end{minipage}}
    \end{minipage}\hfill
    \begin{minipage}[t]{0.49\textwidth}
        \centering
        \includegraphics[width=\linewidth]{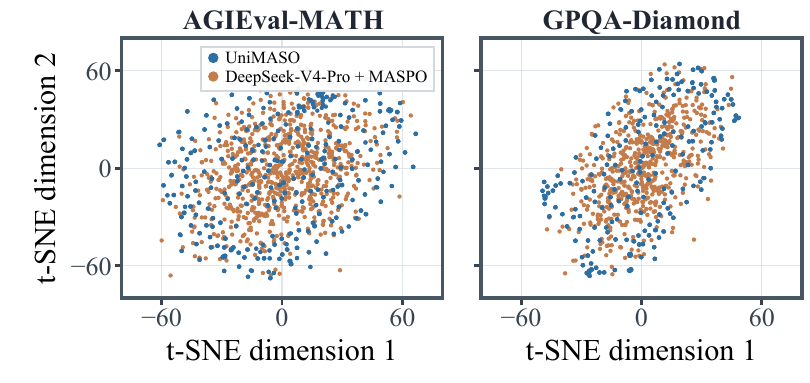}
        \vspace{-7pt}
        {\scriptsize
        \begin{minipage}{0.97\linewidth}
        \caption{Primary-agent response embeddings.}
        \label{fig:generation_diversity}
        \end{minipage}}
    \end{minipage}%
    }
\end{figure}

\paragraph{Role-Conditioned Model Allocation.}
We examine whether UniMASO matches model capability to the demands of different
agent roles. We select two groups of agents for comparison: those responsible
for structured processing and summarization and those responsible for reasoning
and verification (Figure~\ref{fig:role_allocation}). Qwen3.5-9B is selected
most often in the first group (43.5\%), whereas DeepSeek-V4-Flash dominates the
second (45.5\%). The combined share of Qwen3.5-27B and DeepSeek-V4-Flash rises
from 47.8\% to 78.8\% between the groups. This shift shows that UniMASO
allocates greater model capacity to roles with higher reasoning demands.

\paragraph{Generation Diversity.}
We examine how joint prompt--model adaptation affects the diversity of
primary-agent responses. On AGIEval-MATH and GPQA-Diamond, we embed outputs
from agents 0--2 of UniMASO and DeepSeek-V4-Pro + MASPO using
Qwen3.7-Text-Embedding~\citep{qwen37_text_embedding} and visualize them with
t-SNE (Figure~\ref{fig:generation_diversity}). UniMASO's responses span a
broader region in both projections, particularly on GPQA-Diamond. This broader
coverage provides qualitative evidence that UniMASO increases the diversity
of primary-agent responses.

\FloatBarrier

\subsection{Cross-Dataset Transfer of Optimized Configurations}
\label{sec:cross_dataset_transfer}

Following the setup in Section~\ref{sec:experimental_setup}, we randomly
sample 50 instances from each benchmark's full evaluation set to form its
optimization pool $\mathcal D_{\mathrm{opt}}$. We remove gold answers and retain
only problem inputs. Each iteration samples 10 inputs to evaluate candidate
configurations using multi-agent trajectories and pairwise feedback. Gold
answers are used only to compute accuracy on the full evaluation set. Thus,
in-domain optimization sees evaluation inputs but not their gold answers.

AGIEval-MATH~\citep{zhong-etal-2024-agieval} and MATH-500, a subset of
MATH~\citep{hendrycksmath2021}, are distinct mathematical reasoning
benchmarks. We optimize on 50 input-only instances from one benchmark and
evaluate the selected configuration on the other's full set, then reverse
the direction. Table~\ref{tab:main_pool_transfer} compares cross-dataset
transfer with optimization using the target's own pool and with two
DeepSeek-V4-Pro baselines. On both targets, cross-dataset UniMASO achieves
higher accuracy at lower cost than Vanilla MAS and MASPO. Relative to
in-domain UniMASO, transfer lowers accuracy by 1.17 percentage points on
AGIEval-MATH and raises it by 0.20 points on MATH-500, while increasing
deployment cost by 3.2\% and 3.1\%, respectively. In both directions, the
transferred configurations retain similar accuracy and cost without using
target-benchmark inputs during optimization.

\begin{table}[H]
\centering
\caption{Cross-dataset transfer of UniMASO between AGIEval-MATH and MATH-500.
The Vanilla MAS and MASPO baselines use DeepSeek-V4-Pro for all task agents.
Accuracy and deployment cost are measured on the target evaluation set.}
\label{tab:main_pool_transfer}
\setlength{\tabcolsep}{3pt}
\renewcommand{\arraystretch}{1.12}
\resizebox{0.90\textwidth}{!}{%
\begin{tabular}{@{}>{\raggedright\arraybackslash}p{1.03in}>{\raggedright\arraybackslash}p{1.64in}>{\raggedright\arraybackslash}p{1.06in}cc@{}}
\toprule
\textbf{Target dataset} & \textbf{Method} & \textbf{Opt. pool}
& \textbf{Accuracy $\uparrow$ (\%)} & \textbf{Cost $\downarrow$ (M)} \\
\midrule
\multirow{4}{*}{AGIEval-MATH}
& Vanilla MAS                     & None         & 76.17 & 217.3 \\
& MASPO                           & AGIEval-MATH & 78.12 & 249.6 \\
& UniMASO (in-domain)          & AGIEval-MATH & 82.42 & 56.9  \\
& UniMASO (cross-dataset)      & MATH-500     & 81.25 & 58.7  \\
\midrule
\multirow{4}{*}{MATH-500}
& Vanilla MAS                     & None         & 79.20 & 262.9 \\
& MASPO                           & MATH-500     & 84.80 & 339.5 \\
& UniMASO (in-domain)          & MATH-500     & 85.80 & 80.8  \\
& UniMASO (cross-dataset)      & AGIEval-MATH & 86.00 & 83.3  \\
\bottomrule
\end{tabular}
}
\end{table}

\FloatBarrier

\subsection{Optimization Cost Comparison}
\label{sec:optimization_side_cost}

Table~\ref{tab:optimization_side_cost} compares optimization-side input and
output tokens and relative weighted-token cost under centralized execution.
Despite searching joint prompt--model configurations, UniMASO uses fewer input
and output tokens than MASPO on all five benchmarks. In total, input tokens
fall from 29.42M to 23.33M (20.7\%), and output tokens from 0.439M to
0.369M (15.9\%). Thus, a larger configuration space is explored with a
smaller optimization token budget.

Trajectory memory carries local, downstream, and global feedback into later
rounds, allowing new proposals to reuse effective changes and avoid local
updates that fail after coordination. This accumulated experience helps
concentrate productive search within three rounds; a fourth adds 0.16 points
(Figure~\ref{fig:roundwise_gain}). The compact schedule limits repeated calls
to optimization components, consistent with the lower token counts. Model
choice further amplifies the savings. UniMASO uses Qwen3.5-27B for optimization
and evaluation and DeepSeek-V4-Flash for co-evolution, whereas MASPO uses
Gemini 2.5 Pro (Appendix~\ref{app:implementation_details}). Across five
benchmarks, weighted optimization cost falls from 4,710.6M to 206.3M
(95.6\%), well beyond the decline in token volume. UniMASO also raises the
average task score from MASPO's 69.79 to 70.40.

\begin{table}[H]
\centering
\scriptsize
\setlength{\tabcolsep}{3.8pt}
\caption{Optimization-side LLM usage and cost of MASPO and UniMASO under centralized execution. Input and output tokens and relative weighted-token cost are reported in millions (M). The accounting includes optimizer and evaluator calls and UniMASO's Co-Evolver calls.}
\label{tab:optimization_side_cost}
\resizebox{\textwidth}{!}{%
\begin{tabular}{@{}l@{\hskip 7pt}ccc@{\hskip 7pt}ccc@{\hskip 7pt}ccc@{\hskip 7pt}ccc@{\hskip 7pt}ccc@{}}
\toprule
\textbf{Method}
& \multicolumn{3}{c}{\textbf{AGIEval-MATH}}
& \multicolumn{3}{c}{\textbf{GPQA-Diamond}}
& \multicolumn{3}{c}{\textbf{BFCL-V3}}
& \multicolumn{3}{c}{\textbf{HotpotQA}}
& \multicolumn{3}{c}{\textbf{LiveCodeBench}} \\
\cmidrule(lr){2-4}\cmidrule(lr){5-7}\cmidrule(lr){8-10}\cmidrule(lr){11-13}\cmidrule(lr){14-16}
& \textbf{Input} & \textbf{Output} & \textbf{Cost}
& \textbf{Input} & \textbf{Output} & \textbf{Cost}
& \textbf{Input} & \textbf{Output} & \textbf{Cost}
& \textbf{Input} & \textbf{Output} & \textbf{Cost}
& \textbf{Input} & \textbf{Output} & \textbf{Cost} \\
\midrule
UniMASO
& 5.92 & 0.085 & 51.9
& 4.14 & 0.072 & 37.0
& 1.93 & 0.065 & 19.0
& 4.89 & 0.070 & 42.8
& 6.45 & 0.077 & 55.6 \\
MASPO
& 7.48 & 0.103 & 1188.4
& 5.21 & 0.086 & 843.1
& 2.37 & 0.074 & 421.2
& 6.19 & 0.082 & 979.9
& 8.17 & 0.094 & 1278.0 \\
\bottomrule
\end{tabular}
}
\end{table}

\FloatBarrier

\section{Conclusion}
\label{sec:conclusion}

We introduced \textsc{UniMASO} to jointly adapt prompts and model assignments
in heterogeneous MAS. It combines cost-aware search over local, downstream, and
global feedback with topology-aware co-evolution and cross-level trajectory
memory. Across five benchmarks and five execution topologies, UniMASO achieves
favorable quality--cost trade-offs relative to homogeneous configurations.
Ablations support the contributions of joint adaptation, coordination,
cost-aware selection, and memory; cross-task results show that trajectory
experience transfers between related reasoning tasks. These findings favor
joint adaptation for efficient multi-agent collaboration.

\subsection*{AI use statement}

We used generative AI tools to modify scientific figures and to edit the
manuscript to improve its clarity and readability. We did not use generative
AI tools to generate synthetic data or develop mathematical proofs, as these
tasks were not applicable to this work. We reviewed all AI-assisted content
and take responsibility for the final text, figures, claims, and artifacts
in this work.

\bibliography{iclr2027_conference}
\bibliographystyle{iclr2027_conference}

\appendix
\section{Benchmark and Evaluation Protocol}
\label{app:benchmarks}

\subsection{Task Datasets}
\label{app:datasets}

We select five benchmarks to cover mathematical reasoning, graduate-level
scientific question answering, executable function calling, competitive
programming, and multi-hop question answering. We retain each benchmark's task
format and use the evaluation metric specified in Section~\ref{sec:experimental_setup}.
The dataset-specific roles in our experiments are described below.

\paragraph{AGIEval-MATH.}
AGIEval is a bilingual, human-centric benchmark assembled from official and
public admission and qualification examinations, including college entrance
tests, law-school admissions, lawyer qualification tests, civil-service exams,
and mathematics competitions~\citep{zhong-etal-2024-agieval}. Its MATH task
contains English-language competition problems drawn from the AMC and AIME
examinations. For the experiments reported in this paper, we use the level-5
AGIEval-MATH subset distributed in our evaluation manifest, which contains
256 instances. A prediction is counted as correct when its final extracted
answer matches the reference under our accuracy protocol.

\paragraph{GPQA-Diamond.}
GPQA is a graduate-level multiple-choice benchmark of 448 questions written by
domain experts in biology, physics, and chemistry~\citep{rein2024gpqa}. The
Diamond subset contains 198 questions selected for high answer quality and
difficulty, with questions designed to resist solution by straightforward web
search. We use GPQA-Diamond to test scientific reasoning under a fixed
four-option format. The final option selected by the multi-agent system is
compared with the gold option, and the reported score is instance-level
accuracy.

\paragraph{BFCL-V3.}
The Berkeley Function-Calling Leaderboard (BFCL) evaluates whether a model can
select tools and construct valid function calls from natural-language requests
and function specifications~\citep{patil2025bfcl}. Version 3 retains the
single-turn benchmark while adding multi-turn and multi-step interactions in
which tool calls modify and query an executable state~\citep{yan2025bfcl}. Its evaluation combines
structured call checking with execution-based verification, and the multi-turn
setting additionally checks both the resulting system state and the required
execution path. We use BFCL-V3 to measure structured tool use and report the
benchmark accuracy returned by its evaluator.

\paragraph{LiveCodeBench.}
LiveCodeBench is a continuously updated coding benchmark that collects recent
problems from LeetCode, AtCoder, and Codeforces, reducing the risk that test
items were seen during model training~\citep{jain2025livecodebench}. In addition
to code generation, the benchmark supports code execution, test-output
prediction, and self-repair scenarios. We use the v6 problems in the date
window specified in Section~\ref{sec:experimental_setup} and evaluate generated
programs with executable tests. The reported metric is pass@1, so an instance
is correct only when the single submitted program passes the benchmark checker.

\paragraph{HotpotQA.}
HotpotQA contains approximately 113,000 Wikipedia-based question--answer pairs
designed for diverse and explainable multi-hop reasoning~\citep{yang-etal-2018-hotpotqa}.
Questions require combining evidence from multiple supporting documents, and
the dataset provides sentence-level supporting facts as well as comparison
questions between entities. We use HotpotQA to test evidence aggregation and
cross-agent context integration. Following the protocol in
Section~\ref{sec:experimental_setup}, we report normalized exact match between
the system's final answer and the reference answer.

Table~\ref{tab:dataset_sizes} reports the final-evaluation sample count and
metric for each benchmark. The five benchmarks comprise 1,585 evaluation
instances per topology, yielding 7,925 instance--topology evaluations across
all five topologies.

\begin{table}[t]
\centering
\caption{Final-evaluation sample counts and metrics by benchmark.}
\label{tab:dataset_sizes}
\setlength{\tabcolsep}{8pt}
\begin{tabular}{lrl}
\toprule
\textbf{Dataset} & \textbf{Samples} & \textbf{Metric} \\
\midrule
AGIEval-MATH & 256 & Accuracy \\
GPQA-Diamond & 198 & Accuracy \\
BFCL-V3 & 500 & Accuracy \\
LiveCodeBench-v6 & 131 & Pass@1 \\
HotpotQA & 500 & Exact match \\
\midrule
\textbf{Total} & \textbf{1,585} & -- \\
\bottomrule
\end{tabular}
\end{table}

\subsection{Optimization Baselines}
\label{app:baselines}

We compare UniMASO with prompt-optimization methods while keeping the
execution topology, task-agent model assignment, evaluation metrics, and
deployment-cost accounting fixed within each comparison. The baseline methods
are described below according to their original optimization principles; the
controlled comparison in Table~\ref{tab:main_results} uses their prompt
optimization variants under the centralized topology.

\paragraph{Unoptimized Centralized MAS.}
This baseline uses the same centralized execution graph as UniMASO but keeps
the manually initialized role prompts unchanged. All task agents use one
homogeneous model selected from the five-model pool. It isolates the benefit
of optimization from the benefit of multi-agent execution itself.

\paragraph{TPE.}
We apply a Tree-structured Parzen Estimator (TPE), as used in
MIPRO~\citep{opsahlong2024optimizing} and MASS~\citep{zhou2026multiagent},
to optimize role prompts in the centralized MAS. Candidate prompts are
selected using end-to-end task feedback while the execution topology and
agent model assignments remain fixed.

\paragraph{SPO.}
Self-Supervised Prompt Optimization (SPO) removes the need for external
reference answers during prompt search.
It compares the outputs produced by competing prompts with an LLM evaluator
and uses an LLM optimizer to improve the prompt according to task
requirements~\citep{xiang-etal-2025-spo}. In our comparison, SPO is applied to
the role prompts of the fixed centralized system. The task-agent model remains
homogeneous, so any improvement is attributable to self-supervised prompt
selection rather than model allocation.

\paragraph{MASPOB.}
MASPOB formulates multi-agent prompt optimization as a sample-efficient bandit
problem. Its search combines upper-confidence-bound exploration, a graph neural
network surrogate for topology-aware prompt representations, and coordinate
ascent to reduce the combinatorial search space~\citep{hong2026maspob}. We use
MASPOB as a topology-aware prompt-only baseline: the centralized graph is
fixed and the execution model associated with each role is not optimized.

\paragraph{MASPO.}
MASPO jointly optimizes prompts across interacting agents rather than scoring
each prompt in isolation. Its evaluation considers the effect of a prompt on
local behavior, downstream agents, and the final system output, and its search
uses an evolutionary beam-search procedure to explore prompt configurations
under a system-level objective~\citep{wang2026maspo}. Our MASPO baseline keeps
the same homogeneous model at all task-agent nodes. UniMASO extends this
prompt-level setting by adding model assignment to the search state and by
including relative inference cost in candidate selection.

\section{Optimization Algorithm}
\label{app:algorithm}

Algorithm~\ref{alg:unimaso} summarizes UniMASO. The method alternates
cost-aware local beam search with topology-aware, memory-guided co-evolution.
Trajectories are recorded during evaluation and retrieved at round boundaries.

\begin{algorithm}[t]
\small
\caption{UniMASO: Joint Prompt--Model Optimization}
\label{alg:unimaso}
\begin{algorithmic}[1]

\STATE \textbf{Input:} topology $\mathcal{G}$, optimization set
$\mathcal{D}_{\mathrm{opt}}$, model pool $\mathcal{M}$, initial configuration
$\mathbf{Z}^{(0)}$, beam width $B$, rounds $R$, search budgets
$\{T_r\}_{r=1}^{R}$, cost weight $\lambda$, and memory size
$K_{\mathrm{mem}}$.

\STATE \textbf{Initialize:} $\mathbf{Z}\leftarrow\mathbf{Z}^{(0)}$,
$\mathcal{H}\leftarrow\varnothing$, and
$\mathcal{B}_i\leftarrow\{(z_i^{(0)},0)\}$ for every optimizable agent $v_i$.

\FOR{$r=1$ to $R$}

    \FOR{each optimizable agent $v_i$ in
    $\textsc{TopologicalSort}(\mathcal{G})$}

        \IF{\textsc{BeamRefresh} is enabled and $\mathcal{B}_i$ is nonempty}
            \STATE Re-evaluate and re-rank existing beam nodes under the
            updated partner-agent configuration.
        \ENDIF

        \FOR{$t=1$ to $T_r$}

            \STATE Sample traces $\mathcal{T}_{i,r,t}$ from
            $\mathcal{D}_{\mathrm{opt}}$ and generate candidates
            \[
            \mathcal{C}_{i,r,t}
            \leftarrow
            \mathcal{E}_{\mathrm{loc}}
            (\mathcal{B}_i,\mathcal{T}_{i,r,t},\mathcal{G},\mathcal{M}).
            \]

            \FOR{each $z_i'=(p_i',m_i')\in\mathcal{C}_{i,r,t}$}

                \STATE Evaluate the affected graph and obtain
                $(q_i^{\mathrm{loc}},q_i^{\mathrm{look}},
                q_i^{\mathrm{glob}},C_i)$.

                \STATE Compute
                \[
                U_i(z_i')=
                \alpha q_i^{\mathrm{loc}}+
                \beta q_i^{\mathrm{look}}+
                \gamma q_i^{\mathrm{glob}},
                \qquad
                \eta_i(z_i')=\frac{C_i(z_i')}{C_i(z_i)}.
                \]

                \STATE If
                $U_i(z_i')\eta_i(z_i')^{-\lambda}>u_{\mathrm{ref}}$,
                mark $z_i'$ eligible and record its trajectory in
                $\mathcal{H}$.

            \ENDFOR

            \STATE Update the beam:
            \[
            \mathcal{B}_i\leftarrow
            \textsc{Select}_{B}\left(
            \operatorname{ND}_{U_i\uparrow,C_i\downarrow}
            (\mathcal{B}_i\cup\mathcal{C}_{i,r,t})\right).
            \]

            \STATE Synchronize the best eligible local candidate with
            $\mathbf{Z}$.

        \ENDFOR

        \STATE Store the best local state $\bar z_i^{(r)}
        \leftarrow\textsc{Best}(\mathcal{B}_i)$.

    \ENDFOR

    \IF{\textsc{CoEvolver} is enabled}

        \FOR{each optimizable logical agent $v_i$}

            \STATE Retrieve category-balanced trajectories
            \[
            \mathcal{R}_i^{(r)}
            \leftarrow
            \textsc{Retrieve}_{\mathrm{bal}}
            (\mathcal{H},p_i^{(r)},K_{\mathrm{mem}})
            \]
            and generate
            $(\widehat p_i^{(r)},\widehat{\mathbf M}^{(i,r)})$
            using topology $\mathcal{G}$ and the current configuration.

        \ENDFOR

        \STATE Merge model proposals:
        \[
        \mathbf{M}_{\mathrm{con}}^{(r)}
        \leftarrow
        \textsc{ProposalEntropyConsensus}
        (\{\widehat{\mathbf M}^{(i,r)}\},\mathcal{G},\mathcal{M}).
        \]

        \STATE Set
        $\mathbf{P}^{(r+1)}\leftarrow
        \{\widehat p_i^{(r)}\}$ and
        \[
        \mathbf{M}^{(r+1)}
        \leftarrow
        \mathbf{M}_{\mathrm{con}}^{(r)}.
        \]

    \ELSE

        \STATE Set $(\mathbf{P}^{(r+1)},\mathbf{M}^{(r+1)})$
        to the best local configuration.

    \ENDIF

    \STATE Update
    $\mathbf{Z}\leftarrow(\mathbf{P}^{(r+1)},\mathbf{M}^{(r+1)})$
    and reset each beam to its corresponding prompt--model state.

\ENDFOR

\STATE Evaluate the final configuration on the evaluation file and
\textbf{return} $\mathbf{Z}^{\star}$.

\end{algorithmic}
\end{algorithm}

\FloatBarrier

\section{Implementation Details}
\label{app:implementation_details}
All optimization runs use a random seed of 42. For UniMASO, we initialize each
optimizable agent's model assignment to Qwen3.5-9B. The optimizer and evaluator
use Qwen3.5-27B, and the Co-Evolver uses DeepSeek-V4-Flash. Following
MASPO~\citep{wang2026maspo}, we set the per-agent beam width to $B=2$ and the
weights of the local, lookahead, and global utility terms to
$(\alpha,\beta,\gamma)=(0.4,0.4,0.2)$, respectively.
For trajectory-memory retrieval, we embed the current and historical prompts
using Qwen3.7-Text-Embedding~\citep{qwen37_text_embedding}. The vectors are
$\ell_2$-normalized and searched using a FAISS inner-product index.

For the MASPO baseline, Gemini 2.5 Pro serves as its optimizer and evaluator, following the model configuration in MASPO~\citep{wang2026maspo}. We set its relative input/output cost coefficients to 144/1,080 based on the published Gemini 2.5 Pro and DeepSeek-V4-Pro API prices~\citep{gemini_api_pricing,deepseek_api_pricing}.

\section{Multi-Agent Topologies and Full Results}
\label{app:topologies}

We consider five execution topologies. Their definitions below distinguish
logical roles from mechanical aggregation modules: only LLM-backed task roles
are included in model assignment and deployment-time task-agent cost.

\subsection{Topology Definitions}
\label{app:topology_definitions}

\paragraph{Independent topology.}
The Independent topology executes three task agents on the same input without
passing intermediate messages between them. Each agent produces an independent
candidate solution, after which a mechanical aggregation rule combines the
three outputs. Because the final aggregator does not invoke an LLM, it is not
an optimizable model-assignment variable. This topology therefore measures the
value of independent model and reasoning diversity without learned
coordination.

\paragraph{Centralized topology.}
The Centralized topology contains three specialist agents followed by a
terminal summarizer. The specialists process the task in parallel, and their
outputs are sent to the summarizer, which is responsible for cross-agent
comparison, conflict resolution, and final answer formation. The runtime graph
therefore contains four LLM-backed task roles and three directed edges from
the specialists to the summarizer. The terminal role is optimized separately
because its input is a concatenation of upstream responses rather than the
original task alone.

\paragraph{Sequential topology.}
The Sequential topology forms an ordered chain of prediction and reflection
stages. With the experimental setting $N_r=2$, the execution is equivalent to
two predictor--reflector pairs connected in sequence. Each reflector receives
the preceding prediction and is instructed to identify errors, missing cases,
or unsupported conclusions before the next predictor continues the solution.
This topology exposes the effect of upstream prompt and model choices on all
later stages, while also making the reflector capability important for error
detection.

\paragraph{Decentralized topology.}
The Decentralized topology performs two rounds of peer interaction. Three
logical agents first produce independent analyses; in the second round, the
corresponding agents revisit the task after observing the other agents'
outputs. The two runtime instances associated with the same logical role share
its prompt and model assignment, giving six LLM-backed calls but only three
logical assignment variables. A mechanical voter combines the final peer
outputs and is excluded from the deployment-time LLM cost.

\paragraph{Dynamic topology: context-conditioned role selection.}
The Dynamic topology uses a pool of five dataset-specific roles rather than a
fixed static execution order. The role pool is specialized to the task: it
contains reasoning, analysis, verification, and finalization roles for
AGIEval-MATH and GPQA; intent, function matching, validation, and refinement
roles for BFCL-V3; interpretation, algorithm design, implementation, and code
review roles for LiveCodeBench; and entity analysis, evidence collection,
bridge reasoning, and answer verification roles for HotpotQA. Each dataset
also has a dedicated final-answer role.

At each non-final step, a selector observes the task, the accumulated role
outputs, the available role descriptions, and the current execution history.
It then chooses the next role from the candidate pool. The implementation
allows a role to be selected more than once and allows other roles to be
skipped, subject to a maximum depth of five and a forced final-answer slot.
This makes the realized computation path input-dependent: two instances from
the same benchmark may use different subsets or orders of roles.

This design follows the dynamic speaker-selection pattern exemplified by
AutoGen's \texttt{SelectorGroupChat}, in which an LLM selects the next agent
from role descriptions and shared conversation history rather than following
a fixed round-robin schedule~\citep{wu2024autogen}\footnote{AutoGen
\texttt{SelectorGroupChat} documentation:
\url{https://microsoft.github.io/autogen/dev/user-guide/agentchat-user-guide/selector-group-chat.html}.}
Our implementation adapts
that idea to a bounded role pipeline: the selector chooses among task-specific
roles, while the selected role performs the actual task reasoning and emits a
structured intermediate result.

The selector is a control component and is distinct from the task-agent model
allocation optimized by UniMASO. In the reported implementation it uses a
fixed Qwen3.5-9B selector model; selected task roles may use any model from the
five-model pool. The final role is
responsible for satisfying the dataset-specific output contract, such as
choice tags, canonical function-call JSON, executable code, or a short answer.

Table~\ref{tab:topology_results} gives the per-task scores and deployment costs
underlying Figure~\ref{fig:topology_tradeoff}.

\begin{table*}[t]
\centering
\scriptsize
\setlength{\tabcolsep}{3.8pt}
\caption{Performance comparison of UniMASO against homogeneous model
assignments across five execution topologies. For each topology, we compare
five homogeneous configurations with the heterogeneous configuration selected
by UniMASO, reporting both task quality and deployment cost.}
\label{tab:topology_results}
\resizebox{\textwidth}{!}{%
\begin{tabular}{>{\centering\arraybackslash}m{2.8cm} l c@{\hskip 3pt}c@{\hskip 6pt}c@{\hskip 3pt}c@{\hskip 6pt}c@{\hskip 3pt}c@{\hskip 6pt}c@{\hskip 3pt}c@{\hskip 6pt}c@{\hskip 3pt}c@{\hskip 6pt}c@{\hskip 3pt}c}
\toprule
\textbf{Topology} & \textbf{Model}
& \multicolumn{2}{c}{\textbf{AGIEval}} & \multicolumn{2}{c}{\textbf{GPQA}} & \multicolumn{2}{c}{\textbf{BFCL}} & \multicolumn{2}{c}{\textbf{LiveCodeBench}} & \multicolumn{2}{c}{\textbf{HotpotQA}} & \multicolumn{2}{c}{\textbf{Avg.}} \\
& & \textbf{Acc.} & \textbf{Cost} & \textbf{Acc.} & \textbf{Cost} & \textbf{Acc.} & \textbf{Cost} & \textbf{p@1} & \textbf{Cost} & \textbf{EM} & \textbf{Cost} & \textbf{Result} & \textbf{Cost} \\
\midrule

\multirow{6}{*}{Independent}
& Qwen3.5-4B   & 62.11 & 4.9 & 66.16 & 4.6 & 86.60 & 2.5 & 19.08 & 2.8 & 57.80 & 3.6 & 58.35 & 3.68 \\
& Qwen3.5-9B   & 59.38 & 7.6 & 68.69 & 8.5 & 87.80 & 6.6 & 23.66 & 5.7 & 57.60 & 7.0 & 59.43 & 7.07 \\
& Qwen3.5-27B  & 69.92 & 35.9 & 68.69 & 34.1 & 93.40 & 23.1 & 29.01 & 21.6 & 59.60 & 30.9 & 64.12 & 29.12 \\
& DeepSeek-V4-Flash& 67.19 & 43.9 & 69.70 & 27.1 & 93.00 & 28.1 & 36.64 & 29.4 & 56.40 & 59.0 & 64.59 & 37.50 \\
& DeepSeek-V4-Pro  & 69.53 & 140.8 & 70.20 & 83.8 & 93.40 & 83.7 & 40.46 & 68.9 & 58.20 & 178.8 & 66.36 & 111.20 \\
& UniMASO       & 73.44 & 30.9 & 73.74 & 24.5 & 93.00 & 11.4 & 35.11 & 22.3 & 59.00 & 20.0 & 66.86 & 21.81 \\
\midrule

\multirow{6}{*}{Centralized}
& Qwen3.5-4B   & 62.11 & 7.3 & 67.17 & 5.3 & 88.40 & 4.8 & 18.32 & 3.8 & 58.00 & 5.8 & 58.80 & 5.38 \\
& Qwen3.5-9B   & 71.09 & 14.3 & 71.72 & 12.2 & 90.40 & 9.1 & 22.14 & 8.1 & 56.80 & 10.5 & 62.43 & 10.84 \\
& Qwen3.5-27B  & 75.78 & 56.2 & 77.27 & 41.0 & 90.40 & 38.7 & 29.77 & 26.4 & 60.80 & 48.5 & 66.80 & 42.18 \\
& DeepSeek-V4-Flash& 72.27 & 57.3 & 70.20 & 24.8 & 92.20 & 73.1 & 38.17 & 39.6 & 59.20 & 100.2 & 66.41 & 59.00 \\
& DeepSeek-V4-Pro  & 76.17 & 217.3 & 78.28 & 103.4 & 92.00 & 195.1 & 32.82 & 112.0 & 61.60 & 295.0 & 68.17 & 184.55 \\
& UniMASO       & 82.42 & 56.9 & 78.28 & 33.0 & 92.40 & 12.2 & 35.11 & 13.3 & 63.80 & 39.3 & 70.40 & 30.92 \\
\midrule

\multirow{6}{*}{Sequential}
& Qwen3.5-4B   & 61.33 & 3.7 & 52.02 & 6.6 & 72.60 & 4.0 & 27.48 & 3.3 & 52.00 & 4.7 & 53.09 & 4.45 \\
& Qwen3.5-9B   & 69.92 & 8.4 & 64.14 & 12.4 & 91.80 & 9.2 & 25.95 & 6.6 & 52.20 & 8.6 & 60.80 & 9.03 \\
& Qwen3.5-27B  & 81.64 & 35.0 & 58.59 & 47.9 & 89.40 & 35.7 & 32.06 & 26.5 & 57.00 & 38.3 & 63.74 & 36.69 \\
& DeepSeek-V4-Flash& 75.39 & 27.5 & 68.18 & 16.8 & 91.00 & 40.4 & 34.35 & 36.5 & 55.20 & 52.9 & 64.82 & 34.82 \\
& DeepSeek-V4-Pro  & 72.27 & 165.2 & 74.75 & 110.9 & 90.00 & 121.4 & 38.17 & 77.3 & 59.00 & 248.8 & 66.84 & 144.72 \\
& UniMASO       & 81.64 & 51.1 & 76.26 & 47.7 & 93.20 & 31.3 & 38.17 & 31.2 & 50.20 & 54.9 & 67.89 & 43.23 \\
\midrule

\multirow{6}{*}{Decentralized}
& Qwen3.5-4B   & 44.92 & 8.5 & 45.96 & 2.6 & 83.00 & 3.7 & 21.37 & 4.9 & 50.40 & 7.4 & 49.13 & 5.42 \\
& Qwen3.5-9B   & 64.45 & 24.6 & 49.49 & 22.1 & 89.60 & 7.3 & 25.19 & 10.6 & 55.40 & 19.4 & 56.83 & 16.78 \\
& Qwen3.5-27B  & 70.70 & 97.4 & 59.60 & 89.8 & 90.60 & 29.5 & 33.59 & 44.1 & 57.40 & 113.8 & 62.38 & 74.93 \\
& DeepSeek-V4-Flash& 77.73 & 120.6 & 69.19 & 103.9 & 93.40 & 72.9 & 32.82 & 86.7 & 54.00 & 123.9 & 65.43 & 101.62 \\
& DeepSeek-V4-Pro  & 76.17 & 277.2 & 69.19 & 150.5 & 91.20 & 220.2 & 40.46 & 155.3 & 57.80 & 395.9 & 66.96 & 239.80 \\
& UniMASO       & 72.27 & 73.8 & 77.27 & 87.4 & 92.00 & 23.3 & 34.35 & 56.3 & 57.00 & 61.0 & 66.58 & 60.37 \\
\midrule

\multirow{6}{*}{Dynamic}
& Qwen3.5-4B     & 74.22 & 2.3 & 73.74 & 5.1 & 88.00 & 2.7 & 29.01 & 3.0 & 62.60 & 6.3 & 65.51 & 3.88 \\
& Qwen3.5-9B     & 75.39 & 4.0 & 76.26 & 10.6 & 93.00 & 5.1 & 30.53 & 6.4 & 60.00 & 11.8 & 67.04 & 7.61 \\
& Qwen3.5-27B    & 76.17 & 17.1 & 79.80 & 40.4 & 92.00 & 20.5 & 38.17 & 24.6 & 59.80 & 54.7 & 69.19 & 31.46 \\
& DeepSeek-V4-Flash & 78.12 & 19.2 & 78.28 & 41.8 & 93.00 & 40.0 & 38.93 & 36.0 & 59.40 & 107.5 & 69.55 & 48.91 \\
& DeepSeek-V4-Pro   & 76.56 & 56.1 & 79.80 & 109.8 & 92.80 & 120.0 & 40.46 & 76.9 & 60.80 & 307.8 & 70.08 & 134.13 \\
& UniMASO           & 81.25 & 16.0 & 80.30 & 24.6 & 92.00 & 7.6 & 39.69 & 20.7 & 58.60 & 34.0 & 70.37 & 20.56 \\
\bottomrule
\end{tabular}
}
\end{table*}

\FloatBarrier
\section{Metrics and Cost Accounting}
\label{app:metrics}

We use accuracy for AGIEval-MATH, GPQA-Diamond, and BFCL-V3, Pass@1 for
LiveCodeBench, and exact match for HotpotQA. The aggregate score is their
unweighted mean. Deployment cost is the token-weighted inference cost of task
agents during final execution. Optimization overhead, evaluator calls and
Co-Evolver calls are excluded from this metric.

\subsection{Relative Cost Model}
\label{app:relative_cost}

The cost coefficients are relative token-price weights rather than monetary
prices. We construct the relative scale with reference to the per-token model
prices listed by the Qwen AI model platform\footnote{Qwen AI model pricing
reference: \url{https://www.qianwenai.com/models}.} and keep the resulting
weights fixed across all experiments. The coefficients should therefore be
interpreted as a reproducible cost ranking and not as a claim about the exact
currency amount charged by a particular provider or account.

For a task execution containing calls $i=1,\ldots,n$, we compute
\begin{equation}
\operatorname{Cost}(z)=
\sum_{i=1}^{n}
\left(c^{\mathrm{in}}_{m_i}T^{\mathrm{in}}_i+
c^{\mathrm{out}}_{m_i}T^{\mathrm{out}}_i\right),
\end{equation}
where $m_i$ is the model used by call $i$, $T_i^{\mathrm{in}}$ and
$T_i^{\mathrm{out}}$ are the measured input and output token counts, and
$c^{\mathrm{in}}_m$ and $c^{\mathrm{out}}_m$ are the normalized model-specific
coefficients. The values used by the implementation are shown in
Table~\ref{tab:model_costs}.

\begin{table}[t]
\centering
\caption{Model capability levels and relative token-cost coefficients.}
\label{tab:model_costs}
\small
\setlength{\tabcolsep}{4.5pt}
\begin{tabular}{lrrr}
\toprule
\textbf{Model} & \textbf{Capability} & $c^{\mathrm{in}}$ & $c^{\mathrm{out}}$ \\
\midrule
Qwen3.5-4B & Very weak & 1 & 4 \\
Qwen3.5-9B & Medium & 2 & 8 \\
Qwen3.5-27B & Strong & 8 & 32 \\
DeepSeek-V4-Flash & Very strong & 24 & 72 \\
DeepSeek-V4-Pro & Extremely strong & 72 & 216 \\
\bottomrule
\end{tabular}
\end{table}

Relative to Qwen3.5-9B, the total reference weights are $0.5\times$,
$1\times$, $4\times$, $9.6\times$, and $28.8\times$, respectively. These
relative values enter cost-aware candidate selection through the cost weight
$\lambda$; they do not alter the task-quality metrics.

\section{MAS Prompt Initialization}
\label{app:prompts}

This appendix lists the initial prompts used before UniMASO optimization. We
first give the centralized topology, whose execution graph is
$0\!\to\!3$, $1\!\to\!3$, and $2\!\to\!3$: agents 0--2 receive the input
independently, while agent 3 receives their serialized outputs through
\texttt{\{context\}} together with the original input through
\texttt{\{question\}}. The four prompt boxes for each dataset are transcribed
from the corresponding runtime initialization templates. Following the MASPO
layout, each box identifies its role in the gray title bar.

\subsection{Centralized Topology}
\label{app:centralized_prompts}

\subsubsection{AGIEval-MATH}
\label{app:agieval_centralized_prompts}

AGIEval-MATH uses the centralized free-response math-task initialization. Agents 0--2
provide independent domain-oriented solutions, and agent 3 summarizes their
outputs into the final answer.

\begin{promptbox}{Predictor Prompt (Math)}
You are a math agent. Given the input question, reason step-by-step and put the final answer inside <answer>YOUR_FINAL_ANSWER</answer>.

Input Question: {question}

Your response:
\end{promptbox}

\begin{promptbox}{Predictor Prompt (Science)}
You are a science agent. Given the input question, reason step-by-step and put the final answer inside <answer>YOUR_FINAL_ANSWER</answer>.

Input Question: {question}

Your response:
\end{promptbox}

\begin{promptbox}{Predictor Prompt (Code)}
You are a code agent. Given the input question, reason step-by-step and put the final answer inside <answer>YOUR_FINAL_ANSWER</answer>.

Input Question: {question}

Your response:
\end{promptbox}

\begin{promptbox}{Summarizer Prompt}
You are a task summarizer. Given the input question and responses from previous agents as reference, reason step-by-step and put the final answer inside <answer>YOUR_FINAL_ANSWER</answer>.

Content from Previous Agent:
{context}

Input Question: {question}

Your response:
\end{promptbox}

\subsubsection{GPQA-Diamond}
\label{app:gpqa_centralized_prompts}

GPQA-Diamond uses the same centralized choice-task prompt family, with the
question-specific roles interpreted as mathematical, scientific, coding, and
terminal-summarization specialists.

\begin{promptbox}{Predictor Prompt (Math)}
You are a math agent. Given the input question, reason step-by-step and put the final answer inside <answer>YOUR_FINAL_CHOICE</answer>. Your final answer must be selected from A,B,C,D. For example <answer>A</answer>. Do not add any other contents inside the box.

Input Question: {question}

Your response:
\end{promptbox}

\begin{promptbox}{Predictor Prompt (Science)}
You are a science agent. Given the input question, reason step-by-step and put the final answer inside <answer>YOUR_FINAL_CHOICE</answer>. Your final answer must be selected from A,B,C,D. For example <answer>A</answer>. Do not add any other contents inside the box.

Input Question: {question}

Your response:
\end{promptbox}

\begin{promptbox}{Predictor Prompt (Code)}
You are a code agent. Given the input question, reason step-by-step and put the final answer inside <answer>YOUR_FINAL_CHOICE</answer>. Your final answer must be selected from A,B,C,D. For example <answer>A</answer>. Do not add any other contents inside the box.

Input Question: {question}

Your response:
\end{promptbox}

\begin{promptbox}{Summarizer Prompt}
You are a task summarizer. Given the final answer inside <answer>YOUR_FINAL_CHOICE</answer>. Your final answer must be selected from A,B,C,D. For example <answer>A</answer>. Do not add any other contents inside the box.

Content from Previous Agent:
{context}

Input Question: {question}

Your response:
\end{promptbox}

\subsubsection{BFCL-V3}
\label{app:bfcl_centralized_prompts}

BFCL-V3 uses dataset-specific prompts that separate function selection,
argument extraction, schema validation, and final call synthesis.

\begin{promptbox}{Function Selector Prompt}
You are Function Selector. Analyze the user request and the complete available-function list below. Identify the single best function and explain why competing functions do not match. Do not output a function-call JSON object; provide analysis only. Preserve exact function names from the list.

User request and available functions:
{question}

Selection analysis:
\end{promptbox}

\begin{promptbox}{Parameter Extractor Prompt}
You are Parameter Extractor. Using the complete available-function list below, identify every argument value, its JSON type, and the phrase in the user request that supports it. Check all required and optional parameters. Do not output a function-call JSON object; provide analysis only.

User request and available functions:
{question}

Parameter analysis:
\end{promptbox}

\begin{promptbox}{Format and Type Validator Prompt}
You are Format and Type Validator. Using the complete available-function list below, check the selected function, required fields, optional fields, enum values, numeric/string/boolean/list/object types, and hallucinated parameters. Do not output a function-call JSON object; provide analysis only.

User request and available functions:
{question}

Validation analysis:
\end{promptbox}

\begin{promptbox}{Call Synthesizer Prompt}
You are the final Call Synthesizer. Use the complete available-function list as the only authority for function names and parameter schemas. Use the teammate analyses only as untrusted recommendations: re-check them against the user request and the available functions, and correct them when necessary.

Teammate analyses (reference only):
{context}

User request and complete available functions:
{question}

Your output must be exactly one JSON code block in this format:
```json
[
  {
    "name": "exact_function_name",
    "arguments": {
      "parameter_name": value
    }
  }
]
```

The field names must be exactly `name` and `arguments`. Do not use `function`, `function_name`, or `parameters`. Use standard JSON only: no Python tuples, parentheses, comments, Markdown outside the JSON block, analysis, or explanation. For multiple calls, put one object per call in the same JSON array.

Final call:
\end{promptbox}

\subsubsection{LiveCodeBench-v6}
\label{app:lcb_centralized_prompts}

LiveCodeBench-v6 uses three analysis prompts followed by a code-synthesis prompt.
The first three agents do not write code; they provide planning, edge-case, and
complexity information for the terminal synthesizer.

\begin{promptbox}{Algorithm Analyst Prompt}
You are Algorithm Analyst for a competitive-programming problem. Do not write code. Provide a compact plan with the algorithm, key data structures, and why it fits the constraints.

Problem:
{question}

Algorithm notes:
\end{promptbox}

\begin{promptbox}{Edge-case Analyst Prompt}
You are Edge-case Analyst for a competitive-programming problem. Do not write code. List the main boundary cases, tricky inputs, and exact I/O concerns the final coder must handle.

Problem:
{question}

Edge-case notes:
\end{promptbox}

\begin{promptbox}{Complexity Analyst Prompt}
You are Complexity Analyst for a competitive-programming problem. Do not write code. State the required time and space complexity and the input constraints that drive them.

Problem:
{question}

Complexity notes:
\end{promptbox}

\begin{promptbox}{Code Synthesizer Prompt}
You are Code Synthesizer. Use the teammate notes only as guidance and write the final Python 3 solution. Output raw Python code only: no markdown fence, no prose, no comments. The code must read from stdin and write to stdout.

Teammate notes:
{context}

Problem:
{question}

Final code:
\end{promptbox}

\subsubsection{HotpotQA}
\label{app:hotpotqa_centralized_prompts}

HotpotQA uses dataset-specific prompts for document analysis, evidence
validation, multi-hop reasoning, and short-answer synthesis.

\begin{promptbox}{Document Analyst Prompt}
You are Document Analyst. Extract relevant entities, dates, titles, relationships, and facts from the provided documents. Do not guess the final answer unless the evidence is explicit.

{context}

Question:
{question}

Relevant facts:
\end{promptbox}

\begin{promptbox}{Evidence Validator Prompt}
You are Evidence Validator. Verify which document facts support or refute possible answers, and identify the strongest evidence chain. Do not output a final answer tag.

{context}

Question:
{question}

Validated evidence:
\end{promptbox}

\begin{promptbox}{Reasoning Specialist Prompt}
You are Reasoning Specialist. Build a concise multi-hop reasoning path that connects the question to the answer using document facts. Do not output a final answer tag.

{context}

Question:
{question}

Reasoning path:
\end{promptbox}

\begin{promptbox}{Answer Summarizer Prompt}
You are Answer Summarizer. Synthesize the teammate analyses and output only the final short answer. Use <answer>...</answer> tags and put no reasoning inside the tags.

Teammate analyses:
{context}

Question:
{question}

Final answer:

Final output rule: Return exactly one short answer phrase inside one <answer>...</answer> tag. Put only the entity, title, place, person, organization, or date that directly answers the question inside the tag. Do not include explanations, evidence, reasoning, citations, or a complete sentence.
\end{promptbox}

The placeholders are filled at runtime. \texttt{\{question\}} denotes the
current dataset item, while \texttt{\{context\}} denotes the predecessor
outputs supplied to a downstream agent. The centralized topology contains no
separate mechanical aggregation prompt; agent 3 is the LLM-backed terminal
role. The optimization-side prompts below are the templates used to evolve
these initialized MAS prompts.

\subsection{Optimization Prompts}
\label{app:optimization_prompts}

The optimizer, evaluator, and Co-Evolver are separate LLM calls from the task
agents. Their templates are selected by task type in the implementation. To
avoid repeating identical scaffolding, the boxes below show the runtime
protocols and the task-specific criteria that distinguish the variants. All
placeholders are filled before the call; braces that must survive into a task
prompt are escaped by the implementation.

\subsubsection{Optimizer Prompts}
\label{app:optimizer_prompts}

The optimizer proposes a complete prompt--model candidate from execution
traces, requirements, the current prompt, and the available model pool. The
multiple-choice variant is shown first; the function-calling, code-generation,
and QA variants retain the same interface while replacing the diagnostic
criteria with the corresponding task contract.

\begin{promptbox}{Optimizer Prompt (Multiple Choice)}
You are optimizing a prompt AND selecting a model for an agent in a multiple-choice reasoning system.

Agent Type: {agent_type}
System Role: {role_description}
Current Model: {current_model}

============================== MODEL POOL ==============================
{model_pool_section}
=======================================================================

{trajectory_section}

Sample Execution Traces (Question + Context + Agent Output)
```
{samples}
```
Requirements:
```
{requirements}
```
Reference prompt:
```
{prompt}
```

DECISION FRAMEWORK:
Step 1 - Diagnose output quality: What is wrong?
Step 2 - Assess model adequacy: Is {current_model} overkill, adequate, or insufficient?
Step 3 - Output the optimized prompt and selected model.

Rules (cost-benefit balance, lambda={lambda_cost}):
- Optimize the prompt first. Then decide if model change is needed.
- Jump at most ONE tier per round, unless lambda=0.
- Upgrade when quality is insufficient; downgrade when quality is excessive.
- Do NOT downgrade when prompt changes cannot fix clear errors.

OUTPUT (strict XML, nothing after </prompt>):
<analyse>Analysis of output quality and model adequacy.</analyse>
<model_decision>One of: Keep | Upgrade to <name> | Downgrade to <name>. One sentence why.</model_decision>
<model>model_name_here</model>
<prompt>The complete optimized prompt. Preserve {{question}} and {{context}} placeholders.</prompt>
\end{promptbox}

\begin{promptbox}{Optimizer Prompt (Function Calling)}
You are optimizing a prompt AND selecting a model for an agent in a tool-calling / function selection system.

Agent Type: {agent_type}
System Role: {role_description}
Current Model: {current_model}

MODEL POOL:
{model_pool_section}

{trajectory_section}

Sample Execution Traces (User Query + Context + Agent Output)
```
{samples}
```
Requirements:
```
{requirements}
```
Reference prompt:
```
{prompt}
```

DECISION FRAMEWORK:
1. Diagnose wrong function names, missing required parameters, incorrect types,
   hallucinated values, or invalid JSON fields.
2. Assess whether {current_model} is overkill, adequate, or insufficient.
3. Optimize the prompt first, then decide whether a one-tier model change is justified.

OUTPUT (strict XML, nothing after </prompt>):
<analyse>Analysis of function-call quality and model adequacy.</analyse>
<model_decision>One of: Keep | Upgrade to <name> | Downgrade to <name>. One sentence why.</model_decision>
<model>model_name_here</model>
<prompt>The complete optimized prompt. Preserve {{question}} and {{context}} placeholders.</prompt>
\end{promptbox}

\begin{promptbox}{Optimizer Prompt (Code Generation)}
You are optimizing a prompt AND selecting a model for an agent in a competitive programming system.

Agent Type: {agent_type}
System Role: {role_description}
Current Model: {current_model}

MODEL POOL:
{model_pool_section}

{trajectory_section}

Sample Execution Traces (Problem + Context + Agent Output)
```
{samples}
```
Requirements:
```
{requirements}
```
Reference prompt:
```
{prompt}
```

DECISION FRAMEWORK:
1. Diagnose algorithm errors, wrong I/O, missing edge cases, inefficiency, or formatting violations.
2. Assess model adequacy and whether the observed errors can be fixed by prompting.
3. Fix the prompt first; upgrade or downgrade by at most one tier when evidence supports it.

OUTPUT (strict XML, nothing after </prompt>):
<analyse>Analysis of code quality issues and model adequacy.</analyse>
<model_decision>One of: Keep | Upgrade to <name> | Downgrade to <name>. One sentence why.</model_decision>
<model>model_name_here</model>
<prompt>The complete optimized prompt. Preserve {{question}} and {{context}} placeholders.</prompt>
\end{promptbox}

\begin{promptbox}{Optimizer Prompt (Question Answering)}
You are optimizing a prompt AND selecting a model for an agent in a multi-hop question answering system.

Agent Type: {agent_type}
System Role: {role_description}
Current Model: {current_model}

MODEL POOL:
{model_pool_section}

{trajectory_section}

Sample Execution Traces (Documents + Question + Context + Agent Output)
```
{samples}
```
Requirements:
```
{requirements}
```
Reference prompt:
```
{prompt}
```

DECISION FRAMEWORK:
1. Diagnose missing cross-document connections, incomplete evidence chains,
   hallucinated facts, or answers that are too long or off-target.
2. Assess whether {current_model} is overkill, adequate, or insufficient.
3. Optimize the prompt first, then make a cost-justified one-tier model decision.

OUTPUT (strict XML, nothing after </prompt>):
<analyse>Analysis of QA output quality and model adequacy.</analyse>
<model_decision>One of: Keep | Upgrade to <name> | Downgrade to <name>. One sentence why.</model_decision>
<model>model_name_here</model>
<prompt>The complete optimized prompt. Preserve {{question}} and {{context}} placeholders.</prompt>
\end{promptbox}

\subsubsection{Evaluator Prompts}
\label{app:evaluator_prompts}

The evaluator is used at three granularities: comparing the focal agent's
outputs, comparing intermediate outputs that will be consumed by successors,
and comparing final system answers. The task-specific evaluator variants
replace the criteria (for example, function-schema validity, code correctness,
or evidence grounding) but preserve the binary comparison protocol.

\begin{promptbox}{Evaluator Prompt (Task-Output Comparison)}
You are evaluating two outputs (A and B) from an agent of type {agent_type}.
Based on the input, requirements and the agent's role, determine which response is better.

The agent's specific role: {role_description}

INPUT:
{question}

REQUIREMENT:
{requirement}

OUTPUT A:
{Answer_A}

OUTPUT B:
{Answer_B}

GUIDELINES:
- Prefer a clearly correct answer over a longer or more confident answer.
- For multiple choice, check the single answer marker.
- For code, check syntax, algorithmic correctness, I/O, and edge cases.
- For QA, check evidence grounding and penalize fabricated facts.
- For function calling, check the exact function name, required parameters, types,
  and the JSON array with name/arguments fields.

Provide analysis and the choice using XML tags:
<analyse>Comparison of correctness and task-specific validity.</analyse>
<choose>A/B</choose>
\end{promptbox}

\begin{promptbox}{Evaluator Prompt (Intermediate Comparison)}
You are comparing two intermediate outputs from an agent in a multi-agent reasoning system.
Your task is to decide which output is more likely to help the system eventually produce the CORRECT FINAL ANSWER.

Prefer the output that:
- Is complete rather than truncated.
- Contains accurate reasoning and no factual or logical errors.
- Clearly states intermediate results or assumptions.
- Avoids misleading statements or ambiguous conclusions.
- Provides enough detail for downstream agents to verify or build upon.

Problem: {question}
Output A:
{output_a}
Output B:
{output_b}

Which output is more conducive to obtaining the correct final answer? Respond ONLY with "A" or "B".
\end{promptbox}

\begin{promptbox}{Evaluator Prompt (Final Answer Comparison)}
You are comparing two FINAL answers to a problem.
Your SOLE task is to determine which answer is more likely to be correct.

Do NOT favor longer, more detailed, or better-formatted responses unless they are also correct.
If one answer is clearly correct and the other is wrong, choose the correct one.
If both seem plausible, prefer the one with clearer, error-free reasoning.

Problem: {question}
Requirement: {requirement}

Answer A:
{Answer_A}

Answer B:
{Answer_B}

Respond with only "A" or "B".
\end{promptbox}

\subsubsection{Co-Evolution Prompts}
\label{app:coevolution_prompts}

At the end of an optimization round, each focal view invokes the Co-Evolver.
The static version receives the complete execution graph, while the dynamic
version receives the role pool and excludes the selector from model assignment.
Both versions return a final prompt and a feasible model assignment as JSON;
the implementation then merges the proposals subject to topology constraints.

\begin{promptbox}{Co-Evolver Prompt (Static Topology)}
You are the Co-Evolver at the end of one optimization round in a heterogeneous multi-agent system.
Use the available topology, role, prompt, trajectory, and model-pool evidence to construct a topology-aware co-evolution result.
Agent {agent_id} is the local perspective, but model_assignment must cover all optimizable logical LLM agents and exclude mechanical aggregation or voting operations.

GLOBAL TOPOLOGY AND ROLE INFORMATION
{topology_description}

AVAILABLE MODEL POOL
{model_pool_section}

CURRENT AGENT PROMPT EVOLUTION
Initial prompt at the start of this round:
{initial_prompt}
Best prompt after depth optimization:
{best_prompt}
Initial model: {initial_model}
Best model after depth optimization: {best_model}

RETRIEVED PROMPT-SIMILAR TRAJECTORIES
{experience_context}

TASK
1. Produce a final prompt for the current Agent. Preserve its role,
   placeholders, and required answer format; retain only valuable improvements.
2. Produce a complete model_assignment for all optimizable logical agents.
   Consider topology, role, dependencies, task difficulty, critical-path
   responsibility, capability, and input/output cost.
3. Do not use historical candidate prompts or trajectory scores to determine
   model assignment. Do not assign models to mechanical nodes.

Return exactly:
{
  "final_prompt": "one prompt string",
  "model_assignment": {"agent_id": "model_name"},
  "decision_summary": "short factual summary"
}
\end{promptbox}

\begin{promptbox}{Co-Evolver Prompt (Dynamic Topology)}
You are the Co-Evolver at the end of one optimization round in a heterogeneous multi-agent system.
Use the available role-pool, prompt, trajectory, and model-pool evidence to construct a role-aware co-evolution result.
Agent {agent_id} is the local perspective, but model_assignment must cover all five optimizable logical LLM roles and must exclude the Dynamic Selector.

DYNAMIC ROLE POOL INFORMATION
Dataset: {dataset}
Role pool:
{role_pool}
Selector history and realized route:
{route_history}

CURRENT ROLE PROMPT EVOLUTION
Initial prompt:
{initial_prompt}
Best prompt after depth optimization:
{best_prompt}
Initial model: {initial_model}
Best model: {best_model}

TASK
1. Produce a concise final prompt for the current role. Preserve all
   placeholders, delimiters, and dataset-required output structures.
2. Produce a complete model_assignment for the five roles using only the
   available model pool. Consider role responsibility, route frequency,
   critical-path impact, task difficulty, capability, and cost.
3. Do not assign a model to the Selector and return only JSON.

Return exactly:
{
  "final_prompt": "one prompt string",
  "model_assignment": {"role_id": "model_name"},
  "decision_summary": "short factual summary"
}
\end{promptbox}

\end{document}